%% file: apssamp.tex
\documentclass[%
 reprint,
 amsmath,amssymb,
 aps, physrev,
]{revtex4-2}

\usepackage{graphicx}
\usepackage{dcolumn}
\usepackage{bm}
\usepackage{siunitx}
\usepackage[table,xcdraw]{xcolor}
\usepackage{tikz}
\usepackage{color}
\usepackage{xfrac}
\usepackage{array}

\input{results.tex}

\begin{document}

\preprint{APS/Physical Review Accelerators and Beams }

\title{Beam intensity and quality predictions for laser-accelerated ions after capture and transport}

\author{Daniel C. E. Dewitt}
 \email{Contact author: daniel.dewitt@tu-darmstadt.de}
 \affiliation{Institute for Accelerator Science and Electromagnetic Fields (TEMF), Technische Universität Darmstadt, Schlossgartenstrasse 8, 64289 Darmstadt, Germany}

\author{Oliver Boine-Frankenheim}%
 \affiliation{Institute for Accelerator Science and Electromagnetic Fields (TEMF), Technische Universität Darmstadt, Schlossgartenstrasse 8, 64289 Darmstadt, Germany}
\affiliation{GSI Helmholtzzentrum für Schwerionenforschung, Planckstraße 1, 64291, Darmstadt, Germany}

\author{Abel Blažević}%
\affiliation{GSI Helmholtzzentrum für Schwerionenforschung, Planckstraße 1, 64291, Darmstadt, Germany}
\affiliation{%
Helmholtz-Institut Jena, Fröbelstieg 3, 07743, Jena, Germany
}%

\date{\today}

\begin{abstract}
Laser-plasma acceleration produces ultrashort, high-brightness ion beams reaching tens of MeV, yet their large divergence and broad energy spread require dedicated capture elements for beam transport. Using laser-accelerated protons from the GSI PHELIX laser to the LIGHT beamline as a reference, we developed a framework to optimize and assess such combined capture and transport systems, with emphasis on injection into conventional accelerators. 
In addition to our numerical analysis we derive scaling laws linking transmission and chromatic emittance growth to the initial half-opening angle, showing that the present performance is primarily divergence-limited. We also estimate and predict the longitudinal bunch quality and quantify the divergence reduction needed to approach injector-relevant intensities.

\end{abstract}

\maketitle


\section{\label{sec_intro}Introduction}

Target normal sheath acceleration (TNSA) can produce high-current ion beams with particle numbers above $10^{12}$ per shot and very low intrinsic source emittance \cite{cowan0}. These properties make TNSA beams attractive as compact light-ion sources for a range of applications, in particular as injectors for conventional accelerators \cite{busold1} and for medical applications \cite{lhara}. Their practical use is nevertheless limited by the large initial divergence and broad energy spread, which make efficient capture and transport challenging.

A number of studies have addressed the capture and transport of laser-accelerated ion beams using different concepts. These include transport systems based on quadrupoles \cite{chant0,zhu0}, active plasma lenses (APLs) \cite{yang_0,yan_0,yan1}, and pulsed solenoids \cite{kroll1,busold0}. 

At GSI Helmholtzzentrum für Schwerionenforschung, the LIGHT (Laser Ion Generation, Handling and Transport) project \cite{busold1} aims to capture and transport laser-accelerated protons and light ions for injection into the SIS18 (SchwerIonenSynchrotron) via the transfer channel (TK), thereby assessing laser-plasma acceleration as an alternative to conventional ion injectors. Here, we build on the previous transport studies by numerically optimizing a reference LIGHT beamline, consisting of a capture solenoid, a debunching cavity, and a second solenoid, to determine the maximum delivered ion number per shot compatible with transverse and longitudinal acceptance constraints. We show that, despite the low intrinsic source emittance \cite{cowan0}, the large angular divergence and broad energy spread of TNSA beams drive strong emittance growth in both the transverse and longitudinal planes. 

From the derived scaling laws, we identify the needed improvements in the laser parameters on the target.

The paper is structured as follows. After introducing the analytical model for TNSA beams and their transport, we assess the impact of space charge at present and future intensity levels. We then quantify the injected yield of the LIGHT beamline for SIS18 injection, comparing solenoid capture with an active plasma lens (APL). We conclude by discussing scaling laws and upgrade strategies to increase delivered ion numbers and move towards conventional injector performance.

\section{\label{sec_gen_strat}General strategy for capture and transport of laser accelerated ions}

Since TNSA (Target Normal Sheath Acceleration) provides the input beam distribution for all subsequent simulations, this section first summarizes its key characteristics and then introduces the essential beamline elements. The final subsection estimates the impact of space-charge forces. 

The initial TNSA distribution can be characterized as:

\begin{equation}
\label{eq_tnsa_dist}
\Psi(E,\theta) = \frac{d^2N}{dEd\theta}
\end{equation}

with $E$ as the kinetic energy and $\theta$ as the polar emission angle (divergence) of a single particle with respect to the beam axis. We assume that the density-energy distribution $\Psi_{E}(E)$ follows the exponential distribution \cite{mora0}:

\begin{equation}
\label{eq_energy_spread}
\Psi_{E}(E) = \frac{dN}{dE} = \frac{N_{\mathrm{0}}}{\sqrt{2Ek_{B}T}}  \exp{  \left ( -\sqrt\frac{2E}{k_{B}T} \right )}
\end{equation}

with $N_{\mathrm{0}}$ as the total number of particles and $T$ as the hot electron temperature during the TNSA process, both of which are typically determined experimentally and used as fitting parameters. 

The divergence of individual particles emerging from the target we assume as

\begin{equation}
\label{eq_general_divergence}
r'= \theta_{\mathrm{m}}(E) \frac{r}{R_{\mathrm{0}}}
\end{equation}

with the initial radial position $r<R_{\mathrm{0}}$, $\theta_{\mathrm{m}}$ as the half-opening or maximum angle and $r' = dr/dz$ and $R_{\mathrm{0}}$ as the spot size which is significantly smaller than the aperture of the optical elements ($R_{\mathrm{0}} \ll R_{\mathrm{a}}$). The beam is assumed to initially be axially symmetric and to have a homogeneous transverse density.

The half-opening angle $\theta_{\mathrm{m}}$ is typically parametrized by:

\begin{equation}
\label{eq_init_divergence}
\theta_{\mathrm{m}}(E) = \theta_{\mathrm{max}}(1 + a E + b E^2)
\end{equation}

with parameters $a$, $b$, and $\theta_{\mathrm{max}}$ typically derived from radiochromic film measurements \cite{schmitz0, nurnberg0}.

The above model is used to fit experimental or simulation data \cite{schmitz0, lecz0} and will here be used mainly with parameter $\theta_{\mathrm{max}}$ to scale the maximum half-opening angle of the TNSA distributions.

Typically, the energy spread of TNSA beams given by Eq. \ref{eq_energy_spread} ranges from a few MeV up to cutoff energies of several tens of MeV. Using this entire spectrum for simulations or analytical considerations offers little benefit besides additional computational burden which is why we rely on an energy window $\Delta E$ around a reference (kinetic) energy $E_{\mathrm{k}}$.

Moreover, as shown in \cite{hofmann1}, a beamline using a magnetic capture lens inevitably leads to energy filtering at the downstream aperture which is why for the purpose of this study an energy band of \varEnBnd{} is considered. This is approximately the energy window a typical debunching cavity can compensate.

\begin{figure}[ht]
\centering
\includegraphics[width=1\columnwidth]{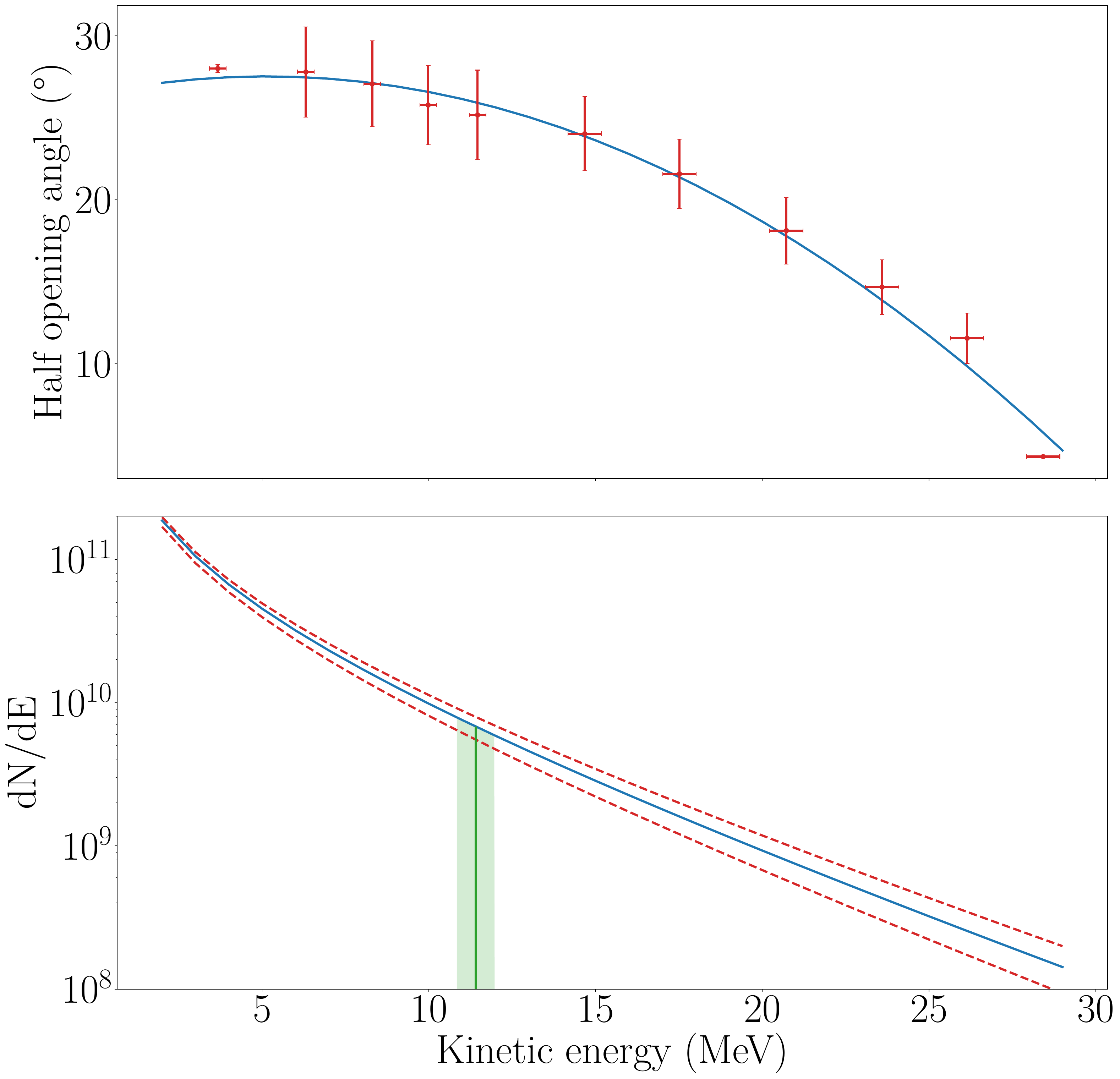}
\caption{Examples for a fitted energy spectrum and half opening angle compared to measurements performed during LIGHT experiments \cite{busold3}. \textbf{Top:} Half-opening angle according to Eq. \ref{eq_init_divergence} (blue) and measurements (red) performed at LIGHT experiments \cite{busold3, metternich0}. Here, $\theta_{\mathrm{max}}$ = \SI{26.5}{\degree}, $a=0.015 \; \mathrm{MeV^{-1}}$ and $b=-1.5 \times 10^{-3} \; \mathrm{MeV^{-2}}$.\textbf{Bottom:} Energy spectrum for a typical TNSA distribution (blue) including an exemplary energy window of \varEnBnd{} (green) and a 95 \; \% confidence band (red) for the estimated dose and hot electron temperature $k_{B}T$.}\label{fig_tnsa}
\end{figure}

Both fitting functions from Eqs. \ref{eq_energy_spread} and \ref{eq_init_divergence} are depicted in Fig. \ref{fig_tnsa} (blue), including the energy window of \varEnBnd{} (green).

Within our model, the TNSA beam envelopes will expand in the transverse plane
as (position of the reference particle $s_{\mathrm{0}}=v_{\mathrm{0}}t$ )
\begin{equation}
\label{eq_tnsa_drift_0}
r_{\mathrm{m}} = \theta_{\mathrm{0}} s_{\mathrm{0}}, \quad r'_{\mathrm{m}} \approx \theta_{\mathrm{0}}
\end{equation}
and longitudinally (half bunch length $z_{\mathrm{m}}$) by
\begin{equation}
\label{eq_tnsa_drift_1}
z_{\mathrm{m}} = z'_{\mathrm{m}}(0) s_{\mathrm{0}}, \quad z'_{\mathrm{m}}(0) \approx \frac{1}{\beta_{\mathrm{0}}^2 \gamma_{\mathrm{0}}^2}\frac{\Delta E}{E_{\mathrm{0}}} 
\end{equation}
with $\theta_{\mathrm{0}} = \theta_{\mathrm{m}}(E_{\mathrm{k}})$, $E_{\mathrm{0}}=\gamma_{\mathrm{0}} mc^2$ as the reference energy and the relativistic parameters $\beta_{\mathrm{0}}$ and $\gamma_{\mathrm{0}}$.

The initial temporal spread at the source is set by the laser–target interaction time, which is in the femtosecond-to-picosecond range and is neglected here ($z_{\mathrm{m}}(0)\approx 0$).

According to Eq. \ref{eq_tnsa_drift_0} the beam radius increases proportional to $s_{\mathrm{0}}$. This leads to the transverse cross section area of the beam increasing proportionally with $s^2_{\mathrm{0}}$. 

For our assumed uniform transverse distribution the transmission $\mathcal{T}$ will thereby decrease by
\begin{equation}
\label{eq_drift_transmission}
\mathcal{T} \propto \frac{1}{s^2_{\mathrm{0}}}
\end{equation}
if the beam passes an aperture at position $s=s_{\mathrm{0}}$.
This makes capture and transport essential for application that are not in direct proximity to the TNSA target.

Longitudinally the beam will expand according to Eq.~\ref{eq_tnsa_drift_1} and a debunching cavity located sufficiently behind the source should reduce $\Delta E$.

Apart from the geometric expansion the normalized transverse emittance also increases linearly depending on the distance $s$ from the source. This effect is mainly driven by the energy spread and divergence of the beam. 
Proportional to $s$ different energy segments in phase space will rotate at a different rate leading to an increase of the normalized transverse emittance.

A detailed discussion is given in \cite{bacci0, floettmann1} and the normalized emittance is estimated as
\begin{equation}
\label{eq_drift_emit}
\varepsilon_{\mathrm{n}} \approx \theta_{\mathrm{0}}^2s_{\mathrm{0}} \frac{\Delta E}{E_{\mathrm{0}}}
\end{equation}
assuming an initially negligible emittance at the source. 

This effect, however, is small compared to the chromatic emittance growth resulting from the first magnetic lens which is discussed in detail in the following section. Here we use $\varepsilon_{\mathrm{n}} \equiv 4 \;\varepsilon_{\mathrm{n,rms}}$, which can be compared more effectively to the transverse acceptance.

In conclusion, transport of TNSA beams requires capture optics to accept the divergent beam and a debunching cavity to reduce the energy spread. Both elements are analyzed in the following section.

\subsection{\label{subsec_capture_lenses}Capture lenses}

So far quadrupoles \cite{hofmann2}, solenoids \cite{kroll0} and active plasma lenses (APLs)\cite{panofsky_0, yang_0, yan_0, yan1} have been implemented as capture elements. Liquid lithium lenses \cite{bayanov0}, similar in function to plasma lenses, are unsuitable in the present low-energy, high-yield TNSA regime since protons experience strong energy loss in lithium (of order \SI{1}{} - \SI{2}{\mega \electronvolt \per \milli \meter}), which would severely reduce the transmitted intensity.

Compared to solenoids, quadrupoles show similar capturing capabilities albeit at lower transmission since at least a quadrupole doublet has to be used \cite{hofmann2}. This leaves solenoids as the primary element of choice for capturing laser accelerated protons with APLs as a possible future alternative currently under development. Similarly, for this study solenoids were considered as capture element with a comparison to APLs being offered in Section \ref{sec_light_optimization}.

Since the first capture element interacts with the full TNSA distribution, its acceptance is a key quantity, as it largely determines the beam delivered to the downstream transport.

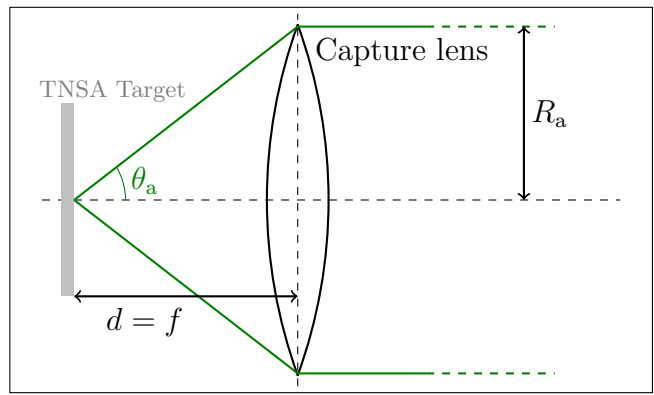
\begin{figure}[ht]
  \centering
  \input{media/lens}
  \caption{Sketch depicting the portion of the TNSA beam (green) passing the lens in with the half-opening angle $\theta_{\mathrm{a}}$. The focal length, essentially the distance to target $d$ is approximated by the distance between target and center of the physical lens.}
  \label{fig_solenoid_beam_transmission}
\end{figure}

Figure \ref{fig_solenoid_beam_transmission} shows the capture of a TNSA beam using a generic lens. In this approximation accepted particles will expand with a divergence angle $\theta < \theta_{\mathrm{a}}$. In this case the acceptance angle can be approximated by $\theta_{\mathrm{a}} \approx R_{\mathrm{a}}/d$ with $d$ being the distance between lens and TNSA source and $R_{\mathrm{a}}$ the aperture of the lens, while assuming a thin lens. Within this approximation a solenoid and a geometrically equivalent APL will have the same acceptance and transmission:

\begin{equation}
\label{eq_en_transmission}
\mathcal{T}_{\mathrm{0}} \approx \left( \frac{\theta_{\mathrm{a}}}{\theta_{\mathrm{0}}} \right)^2
\end{equation}

Here, $\mathcal{T}_{\mathrm{0}}$ is the transmission for a distribution with an energy window $\Delta E$ around the reference kinetic energy of $E_{\mathrm{k}}$ and a half-opening angle of $\theta_{\mathrm{0}} = \theta_{\mathrm{m}}(E_{\mathrm{k}})$.

Another main characteristic of the capture element is the chromaticity, given by $\alpha \equiv \frac{\delta f /f}{\delta E / E}$ which shall here be used to quantify the chromatic aberrations.

For a finite energy window $\Delta E$, the focal length exhibits a corresponding spread $\Delta f$. This chromatic broadening is a general property of magnetic focusing elements; however, it becomes especially severe for TNSA beams because of their large intrinsic energy spread, resulting in enhanced emittance growth \cite{hofmann0}:

\begin{equation}
\label{eq_emittance_growth}
\varepsilon_{\mathrm{n}} = \gamma_0 \beta_0  \alpha f \frac{\Delta E}{E_{\mathrm{k}}} \theta_{}^2
\end{equation}

Here $f$ is the focal length for the reference particle. The emittance growth given by Eq. \ref{eq_emittance_growth} is directly impacted by the chromatic coefficient $\alpha$ which differs between solenoids and APLs.

The difference in chromaticity can be illustrated by the ideal focal-length scalings of the two elements.

The focal length of a solenoid is approximated as:

\begin{equation}
\label{eq_f_sol}
f_{\mathrm{sol}} \approx  \left(  \frac{2p}{qB} \right)^2 \frac{1}{L_{\mathrm{sol}}}
\end{equation}

Here, $B$ is the magnetic field, $q$ the particle charge, $L_{\mathrm{sol}}$ the length of the solenoid and $p$ is the relativistic momentum.

The focal length $f_{\mathrm{APL}}$ of an active plasma lens is given by \cite{yang_0}:

\begin{equation}
\label{eq_f_apl}
f_{\mathrm{APL}}=\frac{1}{\sqrt{G} \sin{(\sqrt{G}L_{\mathrm{APL}})}} \approx \frac{1}{GL_{\mathrm{APL}}}
\end{equation}

with $L_{\mathrm{APL}}$ as the length of the APL and 

\begin{equation}
G\approx \frac{q}{p}\frac{\mu_{\mathrm{0}} I}{2\pi R_{\mathrm{a}}^2}      
\end{equation}
where the plasma radius is $R_{\mathrm{a}}$ and current $I$. For larger beams approaching $r=a$ one can expect nonlinear effects, which we will ignore within this simple estimate.

Due to the $1/p$ or $1/p^2$ dependencies, respectively,  $\alpha_{\mathrm{sol}}\approx 2\alpha_{\mathrm{APL}}$ for the chromatic emittance growth.  

Regarding transmission, solenoids and APLs perform similarly, whereas the lower chromatic factor of an APL reduces chromatic aberrations and hence the emittance ($\varepsilon_{\mathrm{n}} \propto \alpha$).

\subsection{\label{subsec_debunch_cav}Debunching cavity}

As discussed in Section \ref{sec_gen_strat}, a debunching cavity is used to reduce the energy spread of TNSA beams and thereby limit downstream debunching and dispersion. It is placed after a drift that expands the bunch within the approximately linear RF phase interval (e.g., $ \tau_{\mathrm{m}} \lesssim T_{\mathrm{RF}} /6$) \cite{wangler0}; the cavity then performs a longitudinal phase-space rotation to reduce the energy spread.

After a distance of $d_{\mathrm{1}}$ to the source the temporal bunch length is:

\begin{equation}
\label{eq_temporal_growth}
\tau_{\mathrm{m}} = \frac{d_{\mathrm{1}}}{c \beta_0^3 \gamma_0^2} \frac{\Delta E}{E_{\mathrm{0}}}
\end{equation}

For a given voltage, in this case around $qV_{\mathrm{0}} \approx \Delta E$ this leads to the required debunching distance $d_{\mathrm{1}}$ between the TNSA target and cavity kick:

\begin{equation}
\label{eq_comp_approx}
d_{\mathrm{1}} = \frac{\ m c^3 \beta_0^3 \gamma_0^3}{2 \pi qf_{\mathrm{RF}} V_{\mathrm{0}}}
\end{equation}

Here, $f_{\mathrm{RF}}$ is the RF frequency and should not be confused with the focal length $f$. This means that for higher energies either the drift distance $d_{\mathrm{1}}$ before the kick has to be increased or the cavity voltage $V_{\mathrm{0}}$ has to be decreased.

The drifting longitudinal beam distribution is essentially an expanding thin line in $(z,E)$ phase space, with fixed maximum energies.  
The resulting energy width after an ideal cavity kick would therefore be zero. However, due to the initially diverging beam there is a difference in path lengths of $\delta z~\approx~\theta_{\mathrm{a}}^2 d_0 /2$ at the entrance of the capture lens at $d_0$. Due to this local difference in path lengths there is a resulting energy spread in longitudinal phase space given by $\delta E = E' \delta z$ where $E' = dE/dz$ and is approximately $E' \approx 2E/z$. 
This leads to the best achievable total energy spread (after passing the debunching cavity at $s = d_{\mathrm{1}}$) of:

\begin{equation}
\label{eq_drift_long_emit}
\frac{\Delta E_{\mathrm{m}}}{E_{\mathrm{k}}} \approx \theta_{\mathrm{a}}^2 \frac{d_0}{d_1}
\end{equation}

Assuming that the longitudinal phase space ellipse has nearly no tilt after the cavity we can quantify the longitudinal emittance $\varepsilon_z \approx z_{\mathrm{m}} \Delta E_{\mathrm{m}}$ with $z_{\mathrm{m}}$ given by Eq.~\ref{eq_tnsa_drift_1} and $\Delta E_{\mathrm{m}}$ from Eq.~\ref{eq_drift_long_emit}. We chose $z_{\mathrm{m}}(d_1)$ since debunching takes place up to the position of the cavity $d_1$ which leads to:

\begin{equation}
\label{eq_long_emit_growth}
\varepsilon_z = \left ( \frac{1}{\beta_{\mathrm{0}}^2 \gamma_{\mathrm{0}}^2}\frac{\Delta E}{E_{\mathrm{0}}} \right ) d_1 E_{\mathrm{k}}\theta_{\mathrm{}}^2 \frac{d_0}{d_1} = \frac{\Delta E}{\gamma_0(\gamma_0+1)}d_0 \theta_{\mathrm{}}^2
\end{equation}

The initial energy window before the cavity ${\Delta E}/{E_{\mathrm{k}}}$ is constant and for large opening angles the acceptance of the capture element at $s=d_0$ defines $\theta_\mathrm{a}$.

Equations \ref{eq_emittance_growth} and \ref{eq_long_emit_growth} let us estimate the brightness given by

\begin{equation}
\label{eq_brightness}
B_{\mathrm{6D}} = \frac{N}{\varepsilon_{\mathrm{n}}^2 \varepsilon_{\mathrm{z}}} = N \left (  \alpha f d_0 \frac{\Delta E^2}{E_{\mathrm{0}}E_{\mathrm{k}}}  \theta_{\mathrm{}}^6 \right )^{-1}
\end{equation}

assuming an axisymmetric beam. The $\theta^{-6}$ scaling of the brightness again highlights the dominant effect of the initial divergence.   

\subsection{\label{subsec_space_charge}Space-charge effects}

The defocusing space-charge force can potentially limit the captured and transported beam intensity.    
A first estimate of the space charge forces and whether related effects have a relevant impact can be gained by using the envelope equation \cite{reiser0}:
\begin{equation}
\label{eq_env_eq}
r''_m-\frac{K}{r_m}-\frac{\varepsilon_{\mathrm{n}}^2}{\beta^2_{\mathrm{0}} \gamma^2_{\mathrm{0}} r_m^3}=0
\end{equation}
Here, $r_m=x_m=y_m$ is the beam envelope, $r''_m$ denotes $d^2r_m/dz^2$, $K$ is the generalized perveance, and $\varepsilon_{\mathrm{n}}$ is the normalized emittance.
It assumed \cite{busold3} that the TNSA ion beam is charge neutralized by the co-moving cold electrons, when entering the first lens. Space charge defocusing occurs inside and after the first lens, after removal of the electrons.  
Downstream of the capture solenoid, the minimum achievable beam waist is set by the emittance and the generalized perveance $K$. 

\begin{figure}[ht]
  \centering
  \includegraphics[width=1\columnwidth]{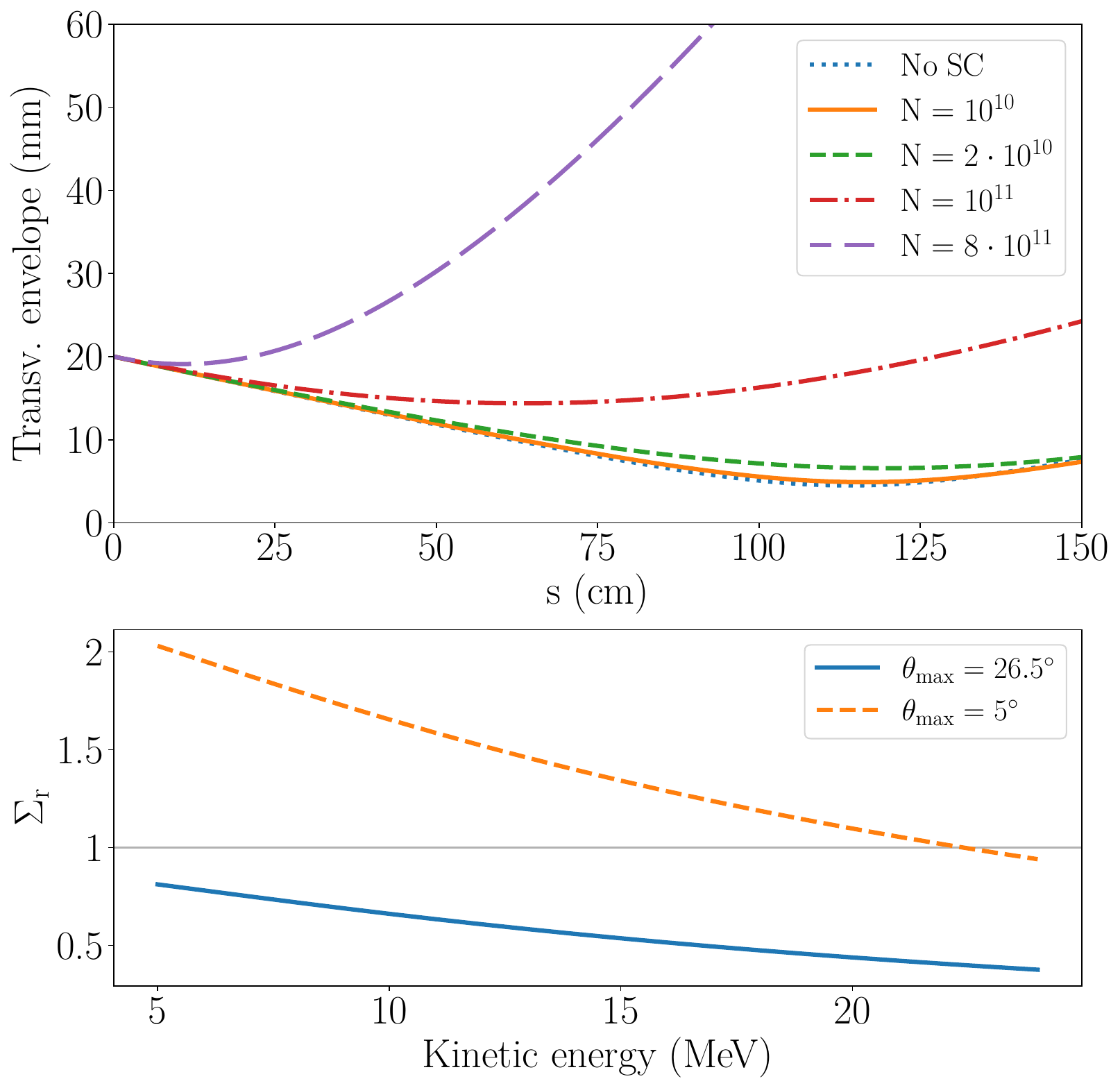}
      \caption{\textbf{Top:} Beam envelopes for different particle numbers, assuming an energy spread of $\varRefEnN \; \mathrm{MeV} \pm \varEnBndN \; \%$ and a normalized emittance of \varEmSolTL{} at the solenoid exit. The envelope obtained without space-charge effects is shown in blue (dotted), while the envelope estimated including the $K_0$ contribution is shown in orange (solid). \textbf{Bottom:} Transverse space-charge parameter for the TNSA spectrum shown in Fig.~\ref{fig_tnsa} for two different values of $\theta_{\mathrm{max}}$.}
  \label{fig_space_charge}
\end{figure}

A simple criterion for when transverse space-charge effects become relevant is given by $K_0/r_{\mathrm{m}} > \varepsilon_{\mathrm{n}}^2/(\beta_{\mathrm{0}}^2 \gamma_{\mathrm{0}}^2 r_{\mathrm{m}}^3)$ in Eq. \ref{eq_env_eq}, at which point the emittance and space-charge contribution to defocusing are comparable. This motivates the definition of the space-charge parameter $\Sigma_{\mathrm{r}}$

\begin{equation}
\label{eq_ch_lim}
\Sigma_{\mathrm{r}} = \frac{K}{K_0} = \frac{K \beta_{\mathrm{0}}^2 \gamma_{\mathrm{0}}^2 r_{\mathrm{m}}^2}{ \varepsilon_{\mathrm{n}}^2} = \frac{qI r_{\mathrm{m}}^2}{ 2 \pi \varepsilon_{\mathrm{n}}^2  \epsilon_0 mc^3\beta_{\mathrm{0}} \gamma_{\mathrm{0}} }
\end{equation}

with the generalized perveance $K$ and the chromatic emittance $\varepsilon_n$ from Eq. \ref{eq_emittance_growth}. Using this parameter, the space-charge dominated regime can be defined as $\Sigma_{\mathrm{r}} > 1$. The corresponding solutions from Eq. \ref{eq_env_eq} are shown in Fig. \ref{fig_space_charge} (top) for different particle numbers. For the chosen reference energy, $\Sigma_{\mathrm{r}} \approx 1$ translates into particle numbers on the order of \varSpChLim{}.

The current $I$ at the solenoid exit can be estimated from Eq. \ref{eq_energy_spread} for a typical TNSA spectrum, using the present reference half-opening angle $\theta_{\mathrm{max}}$ and assuming a corresponding particle loss of 90 \% as estimated from Eq. \ref{eq_en_transmission}. The resulting dependence of $\Sigma_{\mathrm{r}}$ on reference energy is shown in Fig. \ref{fig_space_charge} (bottom). For comparison, the space-charge parameter for a fully transmitted beam is also shown, assuming the same beam intensity with $\theta_{\mathrm{max}} = 5^{\circ}$. 

For the longitudinal plane, we use the envelope equation \cite{reiser0}

\begin{equation}
\label{eq_long_sc_env_eq}
z_{\mathrm{m}}'' = \frac{K_{\mathrm{L}}}{z_{\mathrm{m}}^2} + \frac{\varepsilon_z^2}{z_{\mathrm{m}}^3}
\end{equation}

with the longitudinal perveance

\begin{equation}
\label{eq_long_sc_perveance}
K_{\mathrm{L}} = \frac{3}{2} \frac{gNr_c}{\beta_0^2 \gamma_0^5}
\end{equation}

Here, $r_c$ denotes the classical particle radius. The g-factor is approximated as $g \approx  2z_{\mathrm{m}}/3r_{\mathrm{m}}$ where the bunch radius is taken to be limited by the lens aperture, $r_{\mathrm{m}}~\approx~R_{\mathrm{a}}$. Assuming a bunch defined by the energy selection of the capture lens \cite{hofmann1}, and neglecting the emittance term in Eq. \ref{eq_long_sc_env_eq} for a first estimate, integration from the capture-lens exit to the cavity entrance yields

\begin{equation}
\label{eq_long_sc_debunching_after_sol}
z_{\mathrm{m}}(d_1) \approx z_0 + z_{\mathrm{0}}'(d_1-d_0) + \frac{K_{\mathrm{L}}}{z_0^2} (d_1 - d_0)^2
\end{equation}

The bunch length is $z_0 = z_{\mathrm{m}}(d_0)$ and $z_{\mathrm{0}}'= \Delta E /(\beta_0^2E_0)$. The second term describes the intrinsic debunching of the beam due to the selected energy window $\Delta E$, while the third term estimates the additional debunching caused by space charge between the capture-lens exit at $d_0$ and the cavity entrance at $d_1$. Comparing these two contributions yields a longitudinal space-charge parameter $\Sigma_{\mathrm{z}}$ analogous to Eq. \ref{eq_ch_lim}:

\begin{equation}
\label{eq_long_sc_param}
\Sigma_{\mathrm{z}} = \frac{K_{\mathrm{L}}(d_1 - d_0)}{z_0^2 z_0'}
\end{equation}

For parameters similar to those in Fig. \ref{fig_space_charge} (bottom), this yields $\Sigma_{\mathrm{z}}$ values in the range $10^{-2}$ to $10^{-1}$ at the lowest energies, followed by a steep decline toward higher energies. Longitudinal space-charge effects are therefore negligible. This is mainly due to the already large intrinsic energy spread of TNSA beams, which causes significant debunching even in the absence of space charge.

The results indicate that, for the present beam parameters, the entire TNSA spectrum remains outside the defined space-charge-dominated regime. However, a future reduction in beam divergence at unchanged proton yield would shift the beam into a regime in which transverse defocusing is dominated by space charge. In such a regime, more detailed self-consistent simulations will be required to quantify the impact of space-charge forces on beam transport and to evaluate possible mitigation strategies.

\section{\label{sec_light_optimization}Optimization of the LIGHT setup for maximum transmission}

This section analyzes the potential particle yield of the current system, before turning to scaling laws in the next section. The analytical estimates from Section \ref{sec_gen_strat} are used to validate the simulations presented here, using the LIGHT beamline as a reference model. While the parameters are specific to this study, the analytical and numerical framework is transferable to other beamlines and applications. A comparison between a solenoid and an active plasma lens (APL) is provided in Section \ref{subsec_opt_apl}. Transverse emittances are reported following the convention defined in Section \ref{sec_gen_strat}, $\varepsilon_{\mathrm{n}} \equiv 4 \;\varepsilon_{\mathrm{n,rms}}$ ($2\sigma$-equivalent).

The optimization was performed using a genetic algorithm (GA) \cite{gad0} with the particle tracker ASTRA \cite{floettmann0} via the Python wrapper lume-astra \cite{mayes0}. This workflow is analogous to existing GA-based frameworks such as GIOTTO \cite{conti0}. A Nelder–Mead \cite{nm0} refinement step is applied after the GA to locally fine-tune the solution. The objective function combined maximum transmission with minimal energy spread. Based on the findings in Section \ref{subsec_space_charge}, space-charge effects were neglected for the results presented here. Since ASTRA allows for the inclusion of space charge, the hybrid optimization scheme could in principle be extended to use the GA for global exploration followed by a final local refinement including space-charge effects.

\begin{figure}[!h]
  \centering
  \includegraphics[width=1\columnwidth]{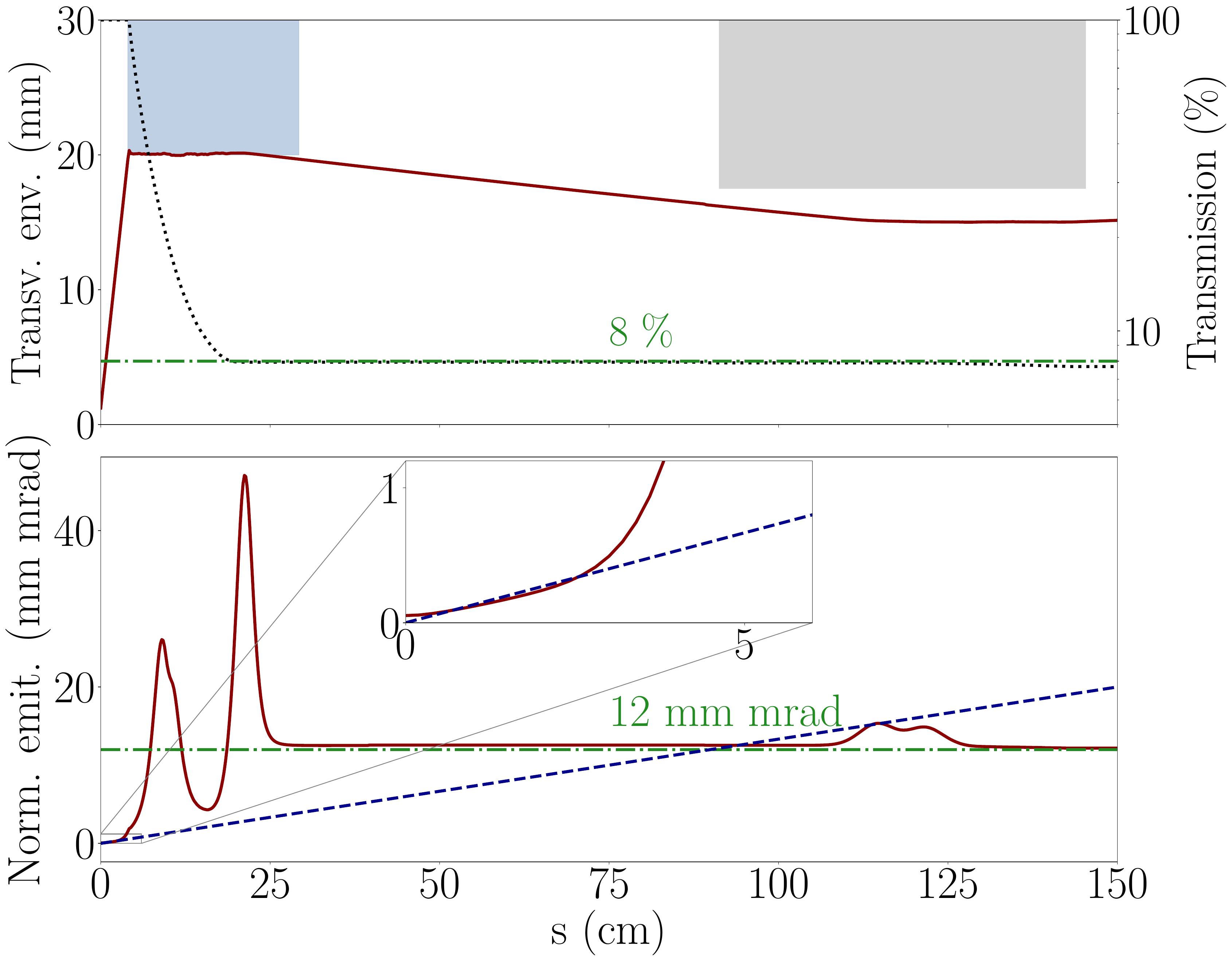}
  \caption{\textbf{Top:} Envelope \textbf{(red, solid)} for an exemplary optimized beamline. The transmission \textbf{(black, dotted)} indicates losses of 90 \% at the capture solenoid. Solenoid and cavity apertures are shown as blue and gray rectangles, respectively. The analytical transmission estimate from Section \ref{subsec_capture_lenses} is shown in \textbf{(green, dash-dotted)}. Assuming axial symmetry of the TNSA beam and beamline elements, only one transverse dimension is shown. \textbf{Bottom:} Emittance for the same beamline configuration \textbf{(red, solid)}, compared with the intrinsic emittance growth in a drift \textbf{(blue, dashed)} estimated by Eq. \ref{eq_drift_emit}. The emittance-growth estimate from Eq.~\ref{eq_emittance_growth} is shown for comparison \textbf{(green, dash-dotted)}. All values are $2\sigma$-equivalent.}
  \label{fig_beam_result}
\end{figure}

Figure \ref{fig_beam_result} shows the transverse envelope and transmission (top) as well as the transverse emittance (bottom) of an optimized beamline for a reference energy of \varRefEn{}. 

The emittance initially grows at a rate well described by Eq. \ref{eq_drift_emit}, and then increases sharply as the beam enters the solenoidal field. This rise is a consequence of ASTRA’s time-based tracking combined with the relatively large energy and temporal spread of the TNSA beam \cite{floettmann1}. Faster particles reach the solenoid field earlier in time than the reference particle, which produces an apparent emittance spike. The effect is neutralized once all particles have left the solenoid at $s \approx 30 \;\mathrm{cm}$, after which the emittance is dominated by the chromatic emittance growth estimated by Eq. \ref{eq_emittance_growth}. A similar effect is observed, though weaker, near the cavity center due to the radial component of the electric field.

Both emittance and transmission are largely dictated by the first capture solenoid acting on the highly divergent TNSA beam. While the emittance is virtually zero at the source, it increases strongly after the capture element, which can be attributed to the beam divergence and energy spread combined with the chromatic aberrations of the magnetic lens. 

\begin{figure}[!h]
  \centering
  \includegraphics[width=1\columnwidth]{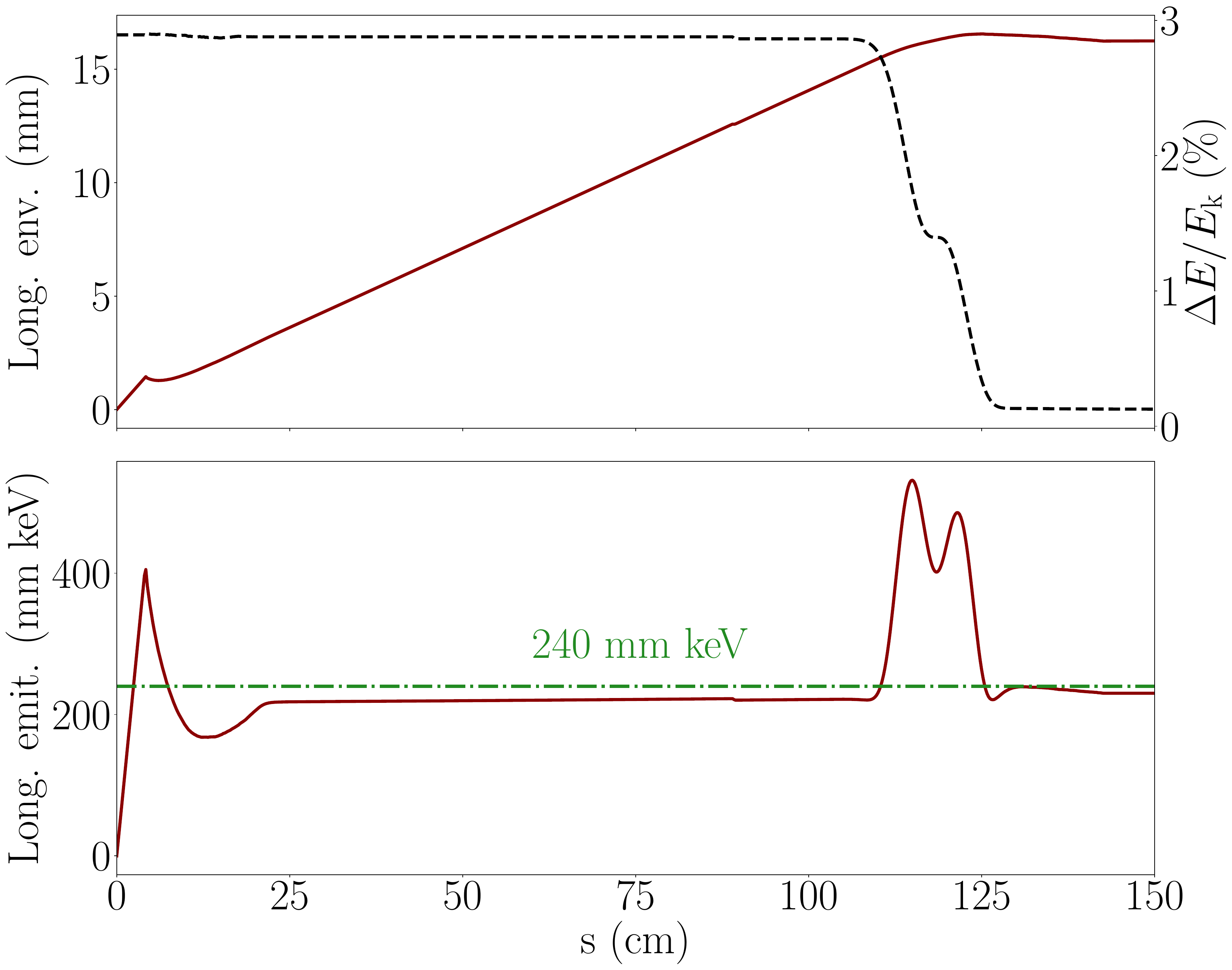}
  \caption{\textbf{Top:} Evolution of the longitudinal envelope \textbf{(red, solid)} and energy spread \textbf{(black, dotted)} relative to the reference energy of \varRefEn{} over $s$. \textbf{Bottom:} Longitudinal emittance along the beamline. The analytical result from Eq. \ref{eq_long_emit_growth} is included as comparison.}
  \label{fig_long_dyn}
\end{figure}

The longitudinal beam evolution is depicted in Fig. \ref{fig_long_dyn}.

Up to the debunching cavity, the energy spread of the beam remains constant (Fig. \ref{fig_long_dyn}, top); according to Eq.~\ref{eq_tnsa_drift_1}, this leads to a linear increase in bunch length. After the cavity kick, the energy spread is reduced and the longitudinal envelope reaches a plateau at around \SI{30}{\milli \meter}. 

The decrease at $s \approx 5 \; \mathrm{cm}$ is caused by losses at the aperture of the capture element.

The longitudinal emittance (Fig.~\ref{fig_long_dyn}, bottom) shows a strong increase up to the capture element as described by Eq. \ref{eq_long_emit_growth}. Assuming a collimated beam, the longitudinal emittance is nearly constant after the capture element at around \SI{200}{\milli \meter \kilo \electronvolt}. The peaks at the position of the cavity ($s \approx 1 \; \mathrm{m}$) are similar evaluation artifacts as discussed for the transverse emittance. 

\begin{figure}[h]
  \centering
  \includegraphics[width=1\columnwidth]{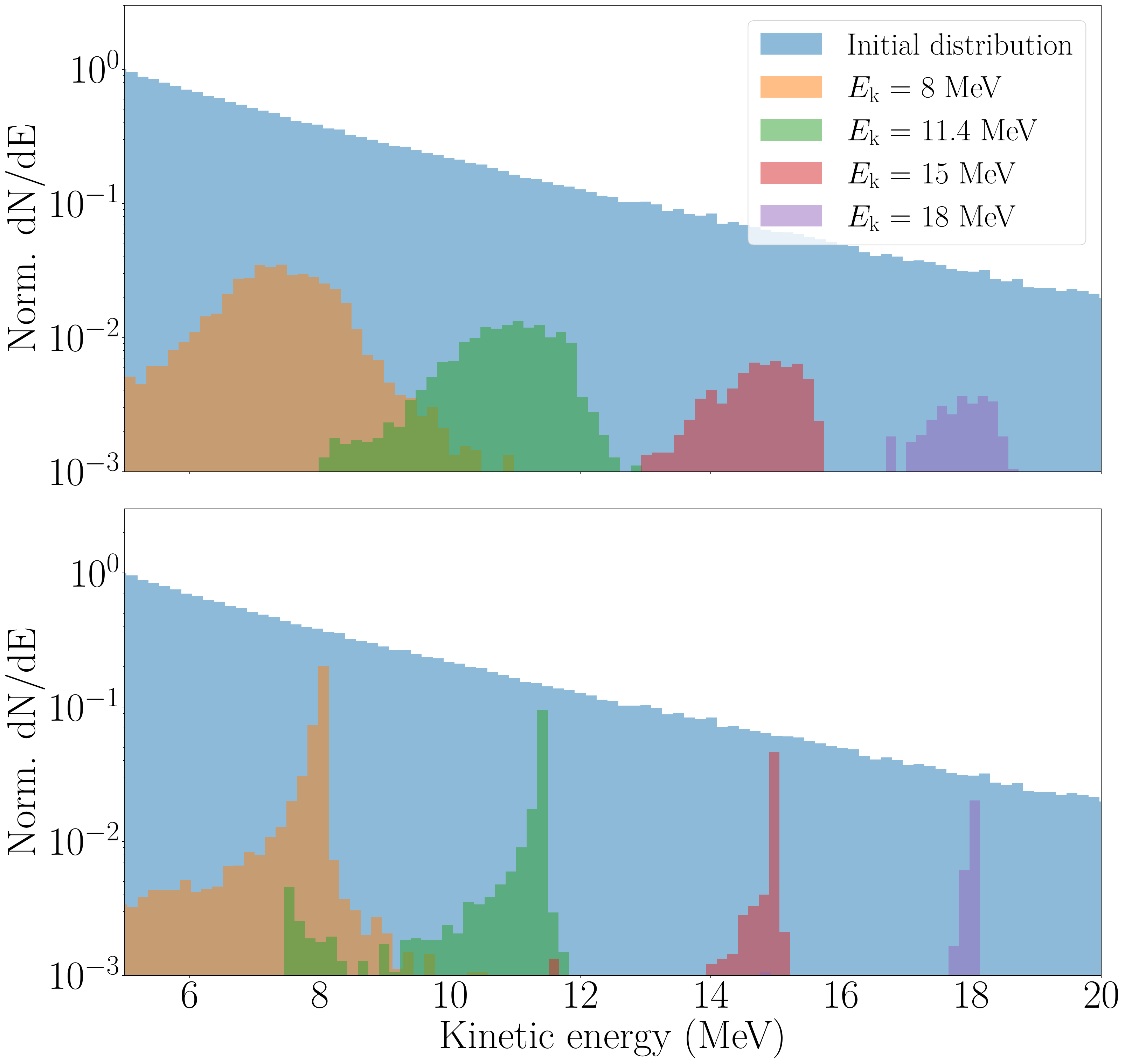}
  \caption{\textbf{Top:} Comparison between the initial TNSA spectrum (blue) and the distribution right before the cavity kick for multiple tracking simulations using different reference energies. The energy selection \cite{hofmann1} is an intrinsic effect of the magnetic capture element and the chromatic aberrations and a subsequent aperture. \textbf{Bottom:} The same distributions as in the \textbf{top} panel but after the debunching cavity kick depicting an increase in particle yield at the specific reference energy of about one order of magnitude.}
  \label{fig_en_histogram}
\end{figure}

Relative to the full TNSA spectrum, the effective transmission is further shaped by the beamline’s intrinsic energy filtering \cite{hofmann1} as illustrated in Fig. \ref{fig_en_histogram}.

Figure \ref{fig_en_histogram} compares the initial TNSA spectrum with the transmitted spectra from simulations for several reference energies. Even without the cavity kick (see Fig. \ref{fig_en_histogram}, top), the energy selection is evident, with the intensity reduced by at least one order of magnitude around the reference energy. The reference energy is set by the magnetic field amplitude of the capture lens. After applying the cavity voltage kick (see Fig. \ref{fig_en_histogram}, bottom), the intensity increases substantially, highlighting the debunching cavity as an essential element of the beamline. The reference beam depicted in Figs. \ref{fig_beam_result} and \ref{fig_long_dyn} is included and depicted in green.

\begin{table}[!h]
  \centering
  \input{media/tab_sis_injection}
    \caption{Reference values and results for injection (inj.) using a solenoid as the capture element. Owing to axial symmetry, the emittances apply to both transverse planes. Highlighted entries indicate the number of protons reaching the TK and SIS18.}
    \label{tab_sis_injection_sol}
\end{table}

The beam parameters at the end of the LIGHT beamline can be compared with the acceptances of the transfer channel (TK) and SIS18 to estimate the number of particles viable for injection. The maximum accepted horizontal emittance of the TK is approximately \SI{20}{\milli\meter\milli\radian}, corresponding to a normalized emittance of about \SI{3}{\milli\meter\milli\radian} at \varRefEn{}; here, we use the more restrictive horizontal acceptance. For typical LIGHT TNSA parameters, the interval $\varRefEnN{} \;\mathrm{MeV} \pm \varEnBndN{} \; \%$ contains about \varRefPartNum{} protons at the source, which defines the density of the initial distribution used in this simulation study.

The results are summarized in Table \ref{tab_sis_injection_sol}. For the present LIGHT configuration, TK can be supplied with \varPartNumEnAcc{} protons at an emittance of \SI{12}{\milli\meter\milli\radian} and an energy spread of approximately $\pm 0.1 \; \%$. After applying the TK acceptance, the proton yield available for SIS18 injection is estimated as \varPartNumTKAcc{}. Relative to conventional injectors such as UNILAC ($>10^{8}$ protons at \SI{36}{\mega\hertz} with \SI{1}{\milli\meter\milli\radian} \cite{barth1}), laser-accelerated sources can reach comparable per-pulse intensities and emittances, albeit at substantially lower repetition rates. Closing the remaining gap in average intensity will require higher-repetition-rate drivers and increased charge per shot. Within the validated analytical–numerical framework developed here, the per-shot yield is primarily limited by the large initial divergence of the TNSA beam and the associated chromatic emittance growth and aperture losses. The following sections therefore use the same framework to (i) assess the potential emittance benefit of an APL capture element and (ii) derive quantitative divergence targets required to approach injector-relevant performance.

\subsection{\label{subsec_opt_apl}Active Plasma Lens as an Alternative Capture Element}

Following Section \ref{subsec_capture_lenses}, active plasma lenses (APLs) can, in principle, outperform solenoidal capture devices for beam transport and focusing mainly due to their lower chromaticity. 
Table \ref{tab_sis_injection_apl} summarizes the results obtained when using the ideal APL as the capture element.

\begin{table}[ht]
  \centering

\input{media/tab_sis_injection_apl}
  \caption{Reference values for injection (inj.) using an active plasma lens (APL) as the capture element. Owing to axial symmetry, the emittances apply to both transverse planes. Highlighted entries indicate the number of protons reaching the TK and SIS18. Acceptance limits and beam parameters are listed in Table \ref{tab_sis_injection_sol}.}
  \label{tab_sis_injection_apl}
\end{table}

The most significant improvement is the reduced emittance, which is about 50 \% lower than for solenoidal capture. This follows from the lower chromaticity of the APL, which behaves more like an axisymmetric quadrupole ($\alpha \approx 1$) than a solenoid ($\alpha \approx 2$), as discussed in Section \ref{subsec_capture_lenses}. This leads to about \varPartNumTKAccAPL{} protons viable for SIS18 injection. 

However, the lower chromaticity also weakens the beamline’s intrinsic energy filtering ($0.2 \; \%$ energy spread after the LIGHT beamline compared to $0.1 \; \%$ for the solenoid - see Tables \ref{tab_sis_injection_apl} and \ref{tab_sis_injection_sol}), increasing the transmitted energy spread and shifting the dominant limitation from the transverse to the longitudinal plane.

The potential advantage of an active plasma lens (APL) relies on an ideal field configuration, i.e., a perfectly linear azimuthal profile ($B_{\varphi}\propto r$). In practice, this ideal is difficult to realize: cooling at the outer wall reduces the plasma conductivity near the boundary, redistributing the current density and steepening the on-axis field gradient ($\partial B_{\varphi}/\partial r$) \cite{tilborg0}. As a result, the magnetic-field profile deviates from linearity, particularly near the outer walls of the APL, introducing strong aberrations across the aperture that can produce ring-shaped beam profiles; this behavior has been observed in simulations and experiments with electron beams \cite{tilborg0}. In our parameter range, these aberrations lead to substantial emittance growth after capture and can increase particle losses by orders of magnitude compared with the ideal case. For this reason, only results for the ideal APL are presented here.

In summary, replacing the capture solenoid would require an APL that closely approaches ideal field characteristics. This has been demonstrated for certain gases (e.g., argon), but only for small radii in the millimeter range \cite{lindstrom0}. Under this assumption an APL as capture element would significantly reduce the chromatically induced transverse emittance.

\section{\label{sec_scal_en_int}Towards higher intensities and energies}

Here we discuss scaling laws to identify the most effective parameters for increasing the delivered charge per pulse and, consequently, the average intensity. Since the TNSA yield scales by $dN/dE \propto I^{1/3}$ \cite{mora0, haines0, kluge0, fuchs0} with laser intensity $I$, reducing the initial divergence at the target \cite{kim0, park0, steinke0, bauer0} is a more promising option. The transmission is expected to scale according to Eq. \ref{eq_en_transmission}, while the chromatic emittance growth (Eq.~\ref{eq_emittance_growth}) scales quadratically with the initial divergence. 
In the following, a solenoid is used as the capture element; the main difference from an APL is the chromaticity (see Section \ref{subsec_capture_lenses}), which affects the emittance and can be rescaled accordingly. 

The first two sections examine how the initial divergence of TNSA beams affects transmission and emittance. Since higher-energy protons are typically less divergent but less abundant, the third section analyzes the energy-dependent trade-off between longitudinal and transverse emittance, and the resulting brightness of the beam. All values are obtained from simulations at the output of an optimized beamline for laser-plasma-accelerated protons.

\subsection{\label{subsec_transmission_scaling}Transmission scaling with initial divergence}

\begin{figure}[ht]
  \centering
  \includegraphics[width=1\columnwidth]{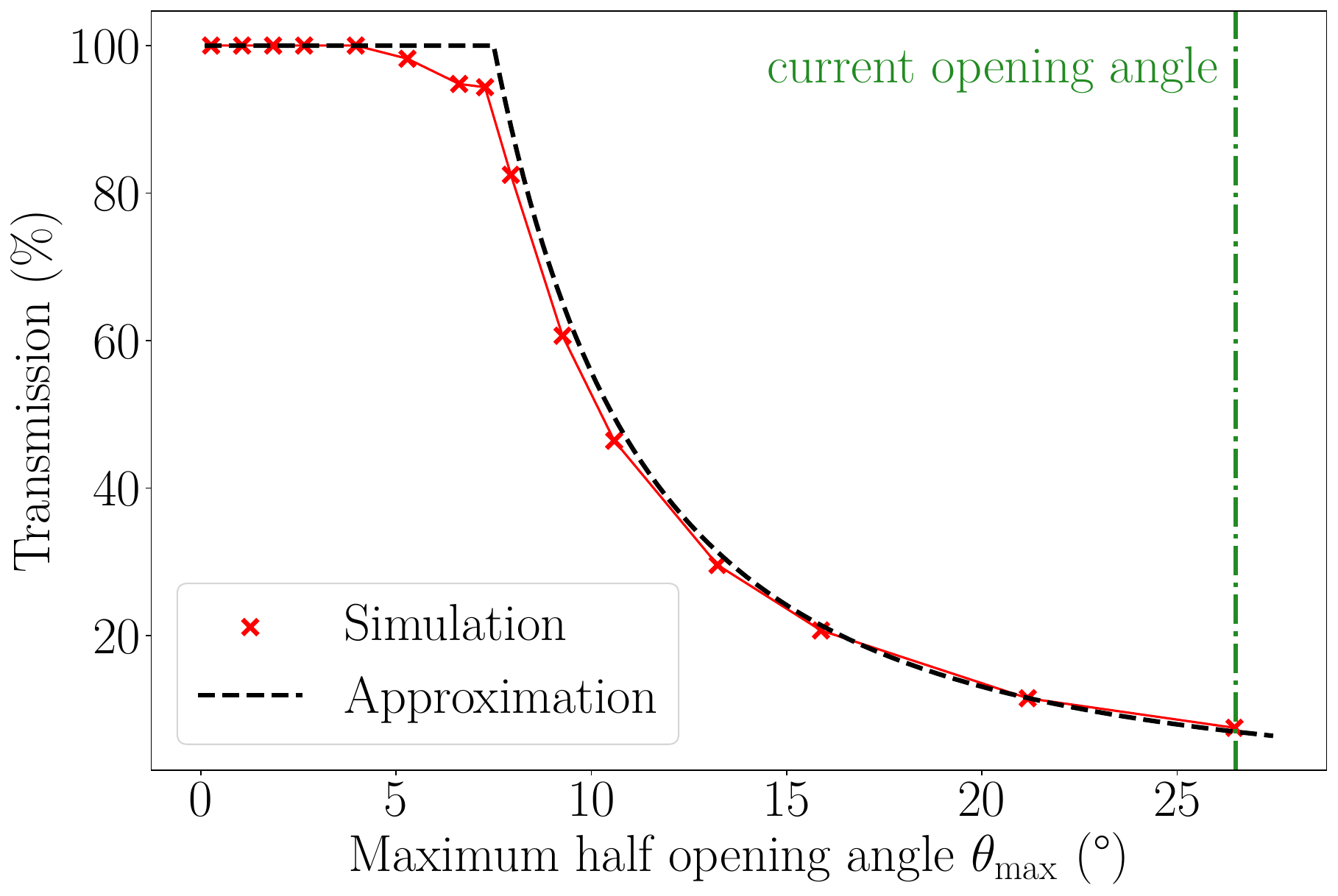}
  \caption{Transmission \textbf{(red, solid)} as a function of the initial half-opening angle $\theta_{\mathrm{max}}$, from the TNSA source to the beamline output. The analytical approximation from Eq. \ref{eq_en_transmission} \textbf{(black, dashed)} is included and shows good agreement with the optimized simulations. The current typical half-opening angle for TNSA beams is shown as reference \textbf{(green, dash-dotted)}.}
  \label{fig_transmission_divergence}
\end{figure}

Analytically the transmission is expected to scale according to Eq. \ref{eq_en_transmission}. Figure \ref{fig_transmission_divergence} shows that the simulations support this scaling, making a reduction of the initial divergence a primary objective.

The slight deviation between simulation and the analytical estimate at $\theta_{\mathrm{max}}\approx \SI{10}{\degree}$ is attributable to additional beamline elements included in the simulation but not captured by Eq. \ref{eq_en_transmission}, which describes only the transmission of the capture element. For half-opening angles below \SI{5}{\degree}, the beam is transmitted through the entire beamline without losses. Achieving this requires reducing the TNSA opening angle by about 80 \%. In this regime, more than $10^{10}$ protons pass the capture element, which - when compared with Fig. \ref{fig_space_charge} - renders space-charge effects non-negligible and motivates appropriate countermeasures.

While full transmission can be achieved at sufficiently small $\theta_{\mathrm{max}}$, the TK acceptance remains well below the emittance delivered by the current beamline (Section~\ref{sec_light_optimization}), motivating a more detailed analysis of this constraint in the next section.

\subsection{\label{subsec_emittance_scaling}Emittance scaling}

\begin{figure}[h]
  \centering
  \includegraphics[width=1\columnwidth]{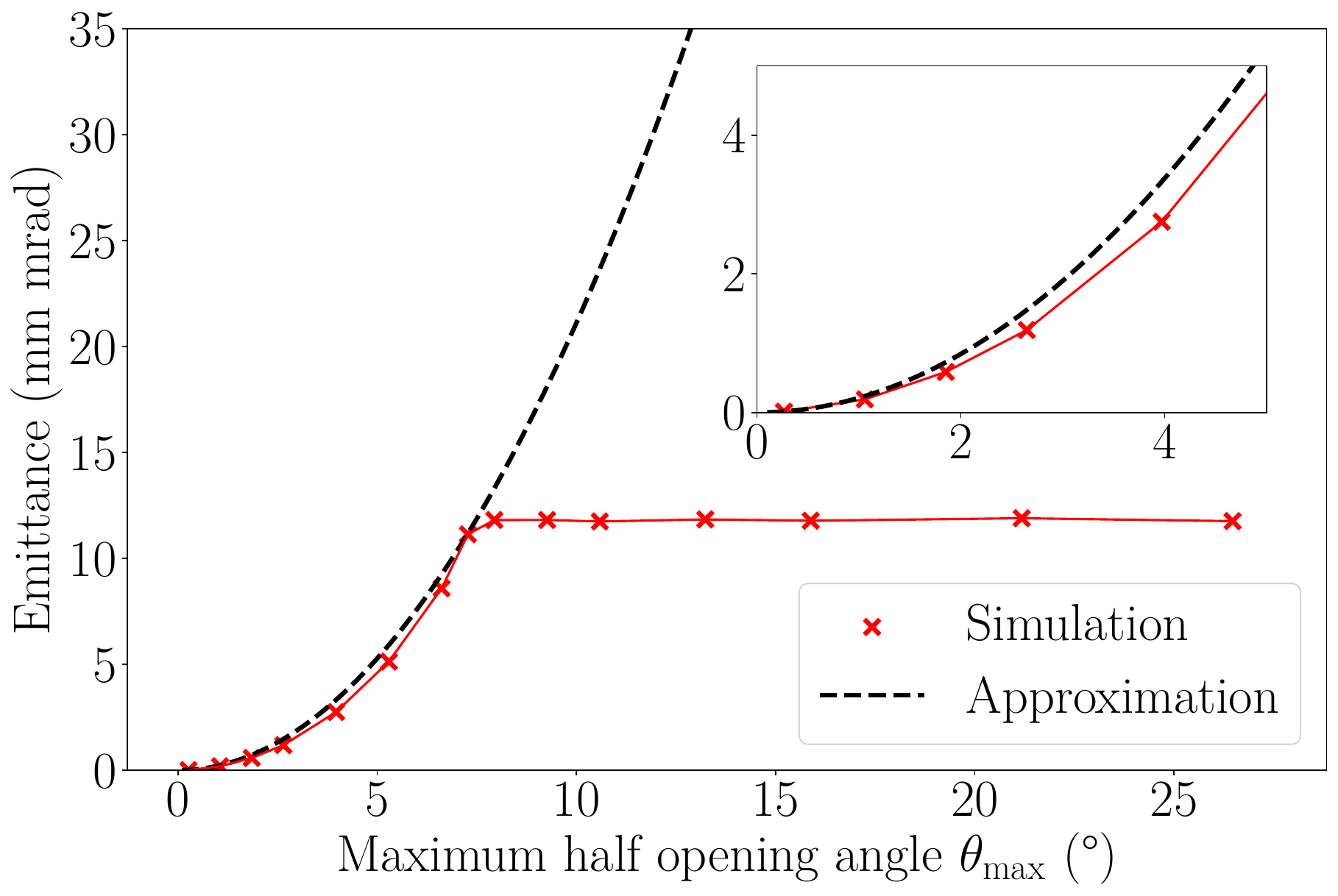}
  \caption{Final emittance \textbf{(red, solid)} as a function of the initial half-opening angle. The analytical estimation \textbf{(black, dashed)} given by Eq. \ref{eq_emittance_growth}, which neglects apertures.}
  \label{fig_emittance_divergence}
\end{figure}

Figure \ref{fig_emittance_divergence} shows the dependence of the normalized emittance on $\theta_{\mathrm{max}}$. For $\theta_{\mathrm{max}} < \theta_{\mathrm{a}}$, the emittance follows Eq.~\ref{eq_emittance_growth}, and the transmission is \SI{100}{\percent}. For larger angles, the emittance becomes limited by the beamline acceptance. This indicates that increasing the radii of beamline elements may increase transmission initially, but at the cost of a larger accepted emittance - effectively shifting losses further downstream rather than eliminating them.
A possible improvement would be the use of an active plasma lens (APL) as the capture element. Since the chromaticity of the magnetic capture lens is the dominant contribution to the emittance growth, an ideal APL would reduce the emittance by roughly a factor of two.
The close-up in Fig. \ref{fig_emittance_divergence} highlights the scaling for $\theta_{\mathrm{max}} < \SI{5}{\degree}$, where the normalized emittance drops below \SI{5}{\milli\meter\milli\radian}. At $\theta_{\mathrm{max}} \approx \SI{4}{\degree}$, the emittance falls below the TK acceptance, such that the full beam within $\Delta E$ is expected to reach the SIS18 injection point. For $\theta_{\mathrm{max}} < \SI{3}{\degree}$, the emittance drops below \SI{1}{\milli\meter\milli\radian}, making the laser-accelerated beam (within $\Delta E$) comparable to UNILAC and, more generally, to conventional injectors. As described in Section ~\ref{subsec_space_charge}, however, the influence of space-charge effects becomes non-negligible at very low $\theta_{\mathrm{max}}$, setting a lower bound on the emittance.

\begin{figure}[h]
  \centering
  \includegraphics[width=1\columnwidth]{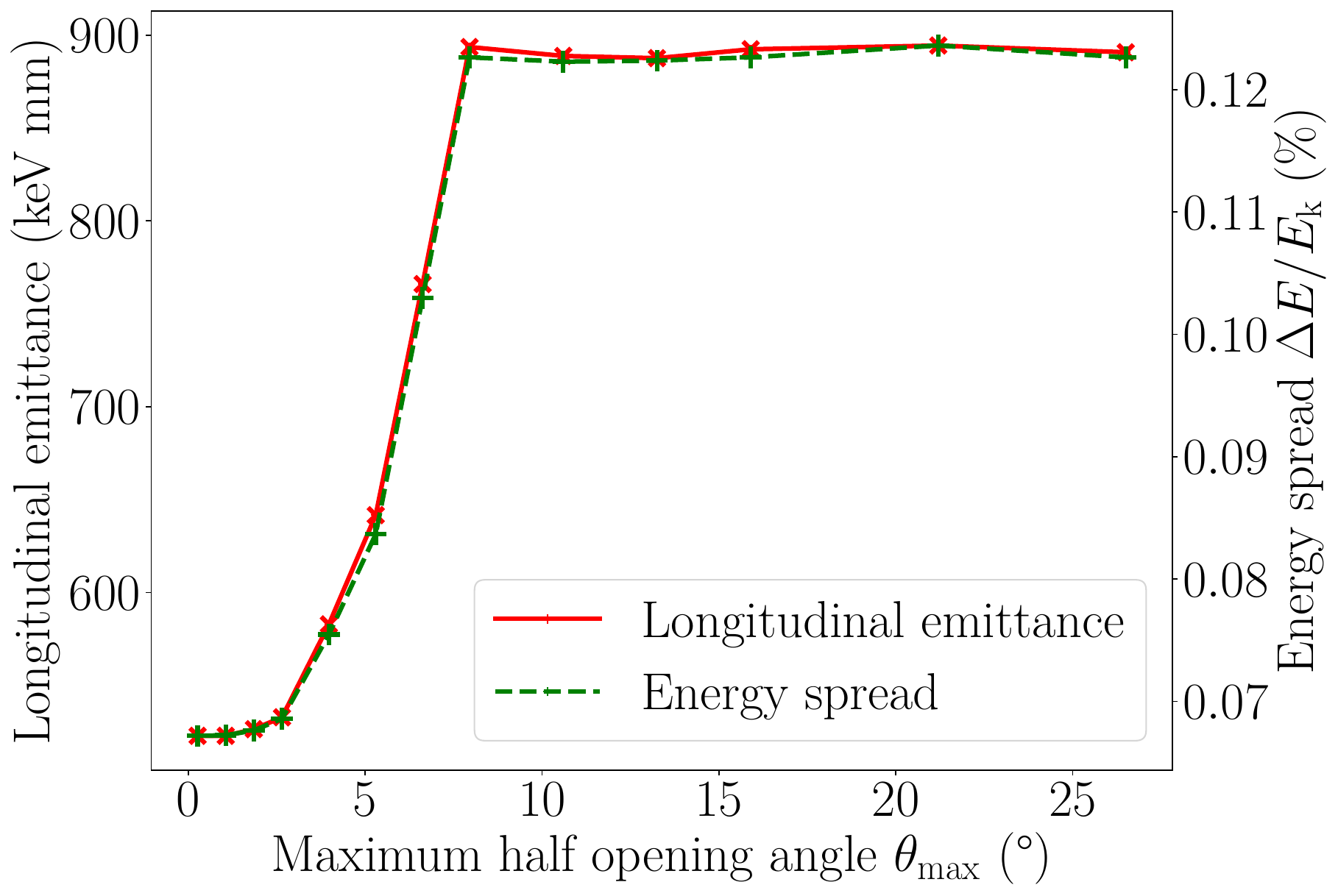}
  \caption{Scaling of the longitudinal emittance and energy spread with the initial half opening angle. Due to the constant distance $d_1$ between source and cavity the longitudinal spread of the beam is also constant, in the absence of space charge. This makes the longitudinal emittance only dependent on the energy spread of the beam.}
  \label{fig_long_emittance_divergence}
\end{figure}

The resulting longitudinal emittance in Fig. \ref{fig_long_emittance_divergence} can be estimated as $\varepsilon_{\mathrm{z}} \approx  z_{\mathrm{m}}\Delta E_{\mathrm{m}}$, with $z_{\mathrm{m}}(d_{\mathrm{1}})$ as the bunch length given by Eq.\ref{eq_tnsa_drift_1}, energy $ \Delta E_{\mathrm{m}}$ given by Eq. \ref{eq_drift_long_emit} and the distance between TNSA source and cavity $d_{\mathrm{1}}$ given by Eq. \ref{eq_comp_approx} which is constant for a given reference energy $E_{\mathrm{k}}$ (see Fig. \ref{fig_long_dyn}). This estimation is valid since ideally, after the cavity kick the longitudinal phase-space ellipse has no tilt. Consequently, the longitudinal emittance follows the same dependence on $\theta_{\mathrm{max}}$ as $\Delta E_{\mathrm{m}} /E_{\mathrm{k}}$. For $\theta_{\mathrm{max}}>\theta_{\mathrm{a}}$, the energy window (and thus $\varepsilon_{\mathrm{z}}$) remains approximately constant, whereas for $\theta_{\mathrm{max}}<\theta_{\mathrm{a}}$ both decrease according to Eq. \ref{eq_energy_spread}.

The remaining energy spread indicates an additional limitation beyond the finite linear phase range of the RF cavity discussed in Section \ref{subsec_debunch_cav}. In the following, we show that nonlinear longitudinal beam dynamics during drift impose a lower bound on the achievable final momentum spread.

\subsection{\label{subsec_optimal_energy}Optimal reference energy}

Equation \ref{eq_init_divergence} shows that the TNSA beam divergence decreases with increasing reference energy. According to Eqs. \ref{eq_emittance_growth} and \ref{eq_drift_long_emit}, this should reduce both the transverse and longitudinal emittance. To isolate these effects from transmission-related losses, a maximum half-opening angle of $\theta_{\mathrm{max}} = 5^{\circ}$ is assumed, corresponding to full transmission, while keeping the particle numbers identical to those used in Table \ref{tab_sis_injection_sol}. The beam is assumed to be axisymmetric. 

\begin{figure}[ht]
  \centering
  \includegraphics[width=1\columnwidth]{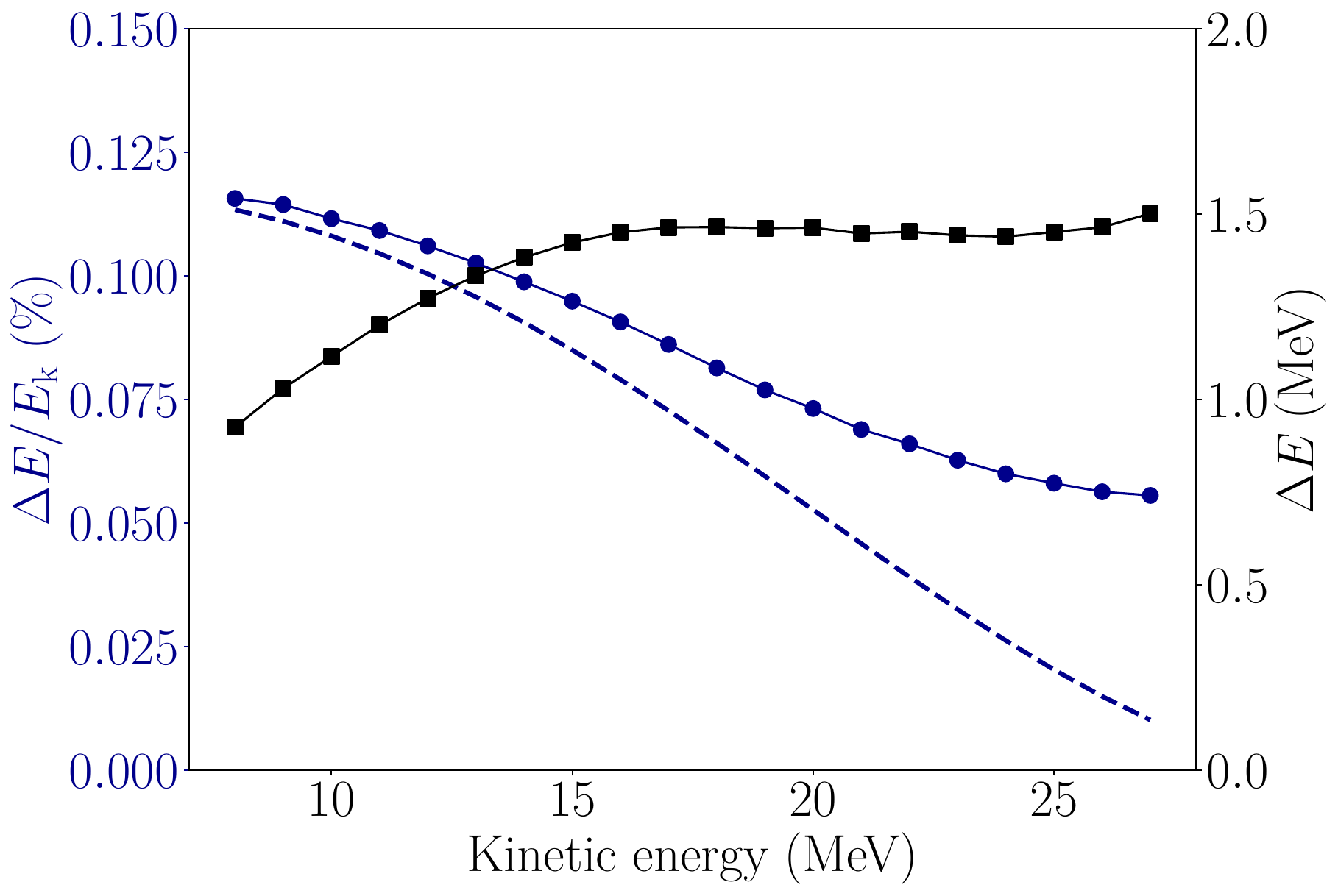}
  \caption{Relative \textbf{(blue)} and absolute \textbf{(black)} energy spread at the end of the optimized beamline as a function of reference energy. The analytical estimate from Eq. \ref{eq_drift_long_emit} is included for comparison.}
  \label{fig_optimal_energy_en_spread}
\end{figure}

Figure \ref{fig_optimal_energy_en_spread} shows the relative and absolute energy spread at the end of the beamline as a function of reference energy. The relative energy spread decreases with increasing reference energy because higher-energy particles in the TNSA spectrum exhibit a lower initial divergence, as described by Eq. \ref{eq_init_divergence}. According to Eq. \ref{eq_drift_long_emit}, the reduced divergence should lead to a relative energy spread that approaches zero, which is not observed in Fig. \ref{fig_optimal_energy_en_spread}.

The deviation between the simulations and Eq.~\ref{eq_drift_long_emit} is based on a linear expansion in $E$. As the divergence decreases, higher-order terms become the limiting factor, as discussed further in Appendix~\ref{app_long_phase_space_curvature}. Similar higher-order limitations of longitudinal phase-space compensation have been discussed for electron beams \cite{zeitler0}. This poses a limit for the achievable energy spread of the beam relative to the initial energy window $\Delta E/E_{\mathrm{k}}$. Since the effective initial energy window is itself determined by the chromatic filtering of the magnetic capture section \cite{hofmann1}, this lower bound reflects a physical limitation of magnetic capture beamlines. The corresponding absolute energy spread, obtained by scaling with the reference energy, approaches an approximately constant value above \SI{15}{\mega\electronvolt}. This results from the compensation between the decreasing relative energy spread and the linear increase in reference energy. 

In conclusion, stricter energy-spread requirements can only be met by reducing the beam divergence or increasing the reference energy. Alternatively, the initial energy window must be narrowed (see Appendix~\ref{app_long_phase_space_curvature}), at the cost of a reduced particle number.

Since the longitudinal emittance can be approximated as $\varepsilon_{\mathrm{z}} \approx z_{\mathrm{m}}\Delta E_{\mathrm{m}}$ and $z_{\mathrm{m}}$ is primarily set by the cavity position, the longitudinal emittance follows the same trend as $\Delta E_{\mathrm{m}}$.

\begin{figure}[ht]
  \centering
  \includegraphics[width=1\columnwidth]{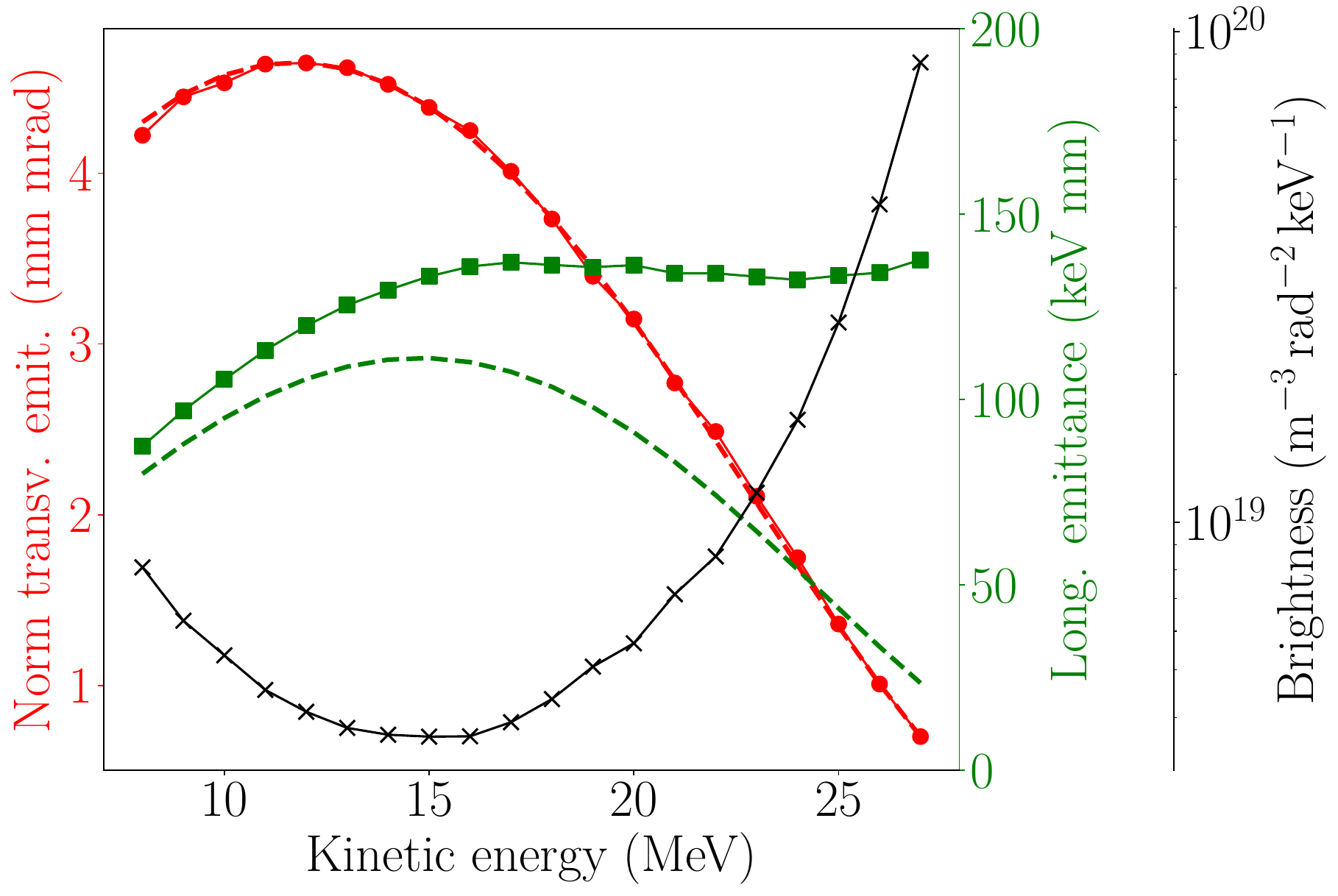}
  \caption{Normalized transverse emittance, longitudinal emittance, and 6D brightness (as defined by Eq. \ref{eq_brightness}) at the end of the beamline as functions of reference energy. The analytical estimates for the longitudinal and transverse emittance from Eqs. \ref{eq_long_emit_growth} \textbf{(green, dashed)} and \ref{eq_emittance_growth} \textbf{(red, dashed)} are included for comparison.}
  \label{fig_optimal_energy}
\end{figure}

The corresponding emittance trends are shown in Fig.~\ref{fig_optimal_energy}. The transverse emittance is primarily determined by the initial beam divergence and therefore decreases strongly with increasing reference energy, in agreement with Eq.~\ref{eq_emittance_growth}. Equation~\ref{eq_long_emit_growth} suggests a similar trend for the longitudinal emittance, but this is not observed in the simulations. The deviation is mainly caused by the relative energy spread shown in Fig.~\ref{fig_optimal_energy_en_spread} and the underlying nonlinearities discussed previously in this section.

Figure~\ref{fig_optimal_energy} further shows the brightness as a function of energy. Although the particle number in TNSA beams decreases approximately exponentially with energy, the brightness increases sharply at higher energies. This behavior is dominated by the strong divergence dependence, $B_{\mathrm{6D}} \propto 1/\theta^6$, and thus reflects the initial opening angle shown in Fig.~\ref{fig_tnsa} and described by Eq.~\ref{eq_init_divergence}. The optimum reference energy is therefore determined by a trade-off between transmitted particle number and beam brightness.

In summary, while beam divergence and higher-order effects set intrinsic bounds, the maximum achievable cavity field imposes the stricter limit on the minimum attainable energy spread. Transversely, the emittance is bounded by its initial value at the TNSA source. This intrinsic emittance is estimated to lie in the µm mrad range \cite{cowan0}, but is expected to increase with laser spot size, while space-charge forces impose an additional lower bound on the emittance. Both limitations will be addressed in future work.

\section{Conclusion}

The present TNSA source-beamline configuration can approach injector-relevant bunch intensities, albeit at repetition rates below those of conventional injectors. Increasing the delivered number of ions per pulse therefore remains a key route to higher average yield. Our analysis indicates that further improvements on the beamline side are limited, with capture-element replacement providing only marginal gains in overall transmission. The analytical scaling laws presented here are consistent with the simulations and enable quantitative target requirements to be formulated for laser-plasma-accelerated protons. In particular, achieving half-opening angles below \SI{5}{\degree} (a divergence reduction of more than 80 \%) appears necessary to reach full transmission through the beamline and to approach injector-relevant operation. Furthermore, for magnetic capture elements chromatic effects lead to a detrimental increase of the projected transverse emittance; for the present reference case, the resulting emittance exceeds the downstream acceptance.

Our numerical particle tracking approach accounts for realistic field configurations and fully nonlinear particle dynamics.
By comparing the obtained results to relatively simple first and second-order approximations for the envelope and emittance evolution we show, that the essential parts of the capture and transport process can be described by reduced models.  

As the key measure we identify the reduction of the initial divergence of TNSA beam. Lower divergence can be pursued through several approaches, including tailored target geometries and an increased laser focal spot size. Higher laser intensities may also increase the proton yield in absolute terms; however, the scaling is weak (sub-linear), such that large intensity increases yield only modest gains. Given the importance of operational stability and high repetition rates, enlarging the focal spot appears to be the most promising route and warrants further investigation. 

In addition, extending the studies to heavier ion species is important, since their divergence and charge-state distribution can substantially modify both capture and downstream injection requirements.

\section*{Data Availability}

The optimization code can be made available upon reasonable request from the authors.

\begin{acknowledgments}
The authors gratefully acknowledge the support of the LIGHT collaboration at GSI Helmholtzzentrum für Schwerionenforschung.

This work is supported by the Deutsche Forschungsgemeinschaft (DFG, German Research Foundation) – Project-ID 499256822 – GRK 2891 'Nuclear Photonics'.
\end{acknowledgments}

\section*{Author Contributions}

\textbf{Daniel Dewitt:} Conceptualization, Methodology, Software, Formal analysis, Investigation, Visualization, Writing – original draft, Writing – review \& editing. \textbf{Oliver Boine-Frankenheim:} Conceptualization, Methodology, Formal analysis, Supervision, Funding acquisition, Writing – review \& editing. \textbf{Abel Blažević:} Resources, Validation, Writing – review \& editing.

\newpage

\appendix

\section{\label{app_long_phase_space_curvature}Longitudinal phase space curvature in TNSA beams}

In the absence of transverse divergence, the linear evolution of the longitudinal phase space can be described as a tilt of an infinitesimally thin line. To estimate the contribution of second-order effects, we consider the drift time of a single particle

\begin{equation}
\label{eq_app_time_drift}
t(E) = \frac{d}{\beta (E)c}
\end{equation}

for a drift distance $d$. A second-order expansion about the energy $E_0 = \gamma_0 mc^2$ yields

\begin{equation}
\label{eq_app_time_ex}
t(E) = t_0+ \xi_1 \Delta E + \frac{\xi_2}{2} \Delta E^2
\end{equation}

with the derivatives:

\begin{equation}
\label{eq_app_derivative_0}
\xi_1 = \frac{dt}{dE}\Bigr|_{E_0} = \frac{d}{c \beta_0^3 \gamma_0^2 E_0}
\end{equation}

\begin{equation}
\label{eq_app_derivative_1}
\xi_2 = \frac{d^2t}{dE^2}\Bigr|_{E_0} = \frac{3d}{c \beta_0^5 \gamma_0^2 E_0^2}
\end{equation}

We estimate the energy $\Delta E_{\mathrm{m}}$ after the cavity kick as

\begin{equation}
\label{eq_app_cavity_kick}
\Delta E_{\mathrm{m}} = \Delta E - k \left ( \xi_1 \Delta E + \frac{\xi_2}{2} \Delta E^2 \right )
\end{equation}

with $\Delta E$ as the initial energy difference before the cavity kick. For an ideal cavity that compensates the linear term in Eq. \ref{eq_app_cavity_kick} one obtains $k = 1/\xi_1$, and the quadratic residual becomes
\begin{equation}
\label{eq_app_residual}
\frac{\xi_2}{2 \xi_1} \Delta E^2 = \frac{3}{2 \beta_0^2} \frac{\Delta E^2}{E_0}
\end{equation}

Related to the reference kinetic energy $E_{\mathrm{k}}$ this simplifies to:

\begin{equation}
\label{eq_app_residual_kinetic}
\frac{\Delta E_{\mathrm{m, res}}}{E_{\mathrm{k}}} = \frac{3 \gamma_0}{2 (\gamma_0 +1)}  \left (\frac{\Delta E}{E_{\mathrm{k}}} \right )^2  
\end{equation}

This shows that, even under the paraxial assumption, nonlinear contributions to the longitudinal phase space produce a residual energy spread that cannot be removed by an ideal cavity. 

\begin{figure}[ht]
  \centering
  \includegraphics[width=1\columnwidth]{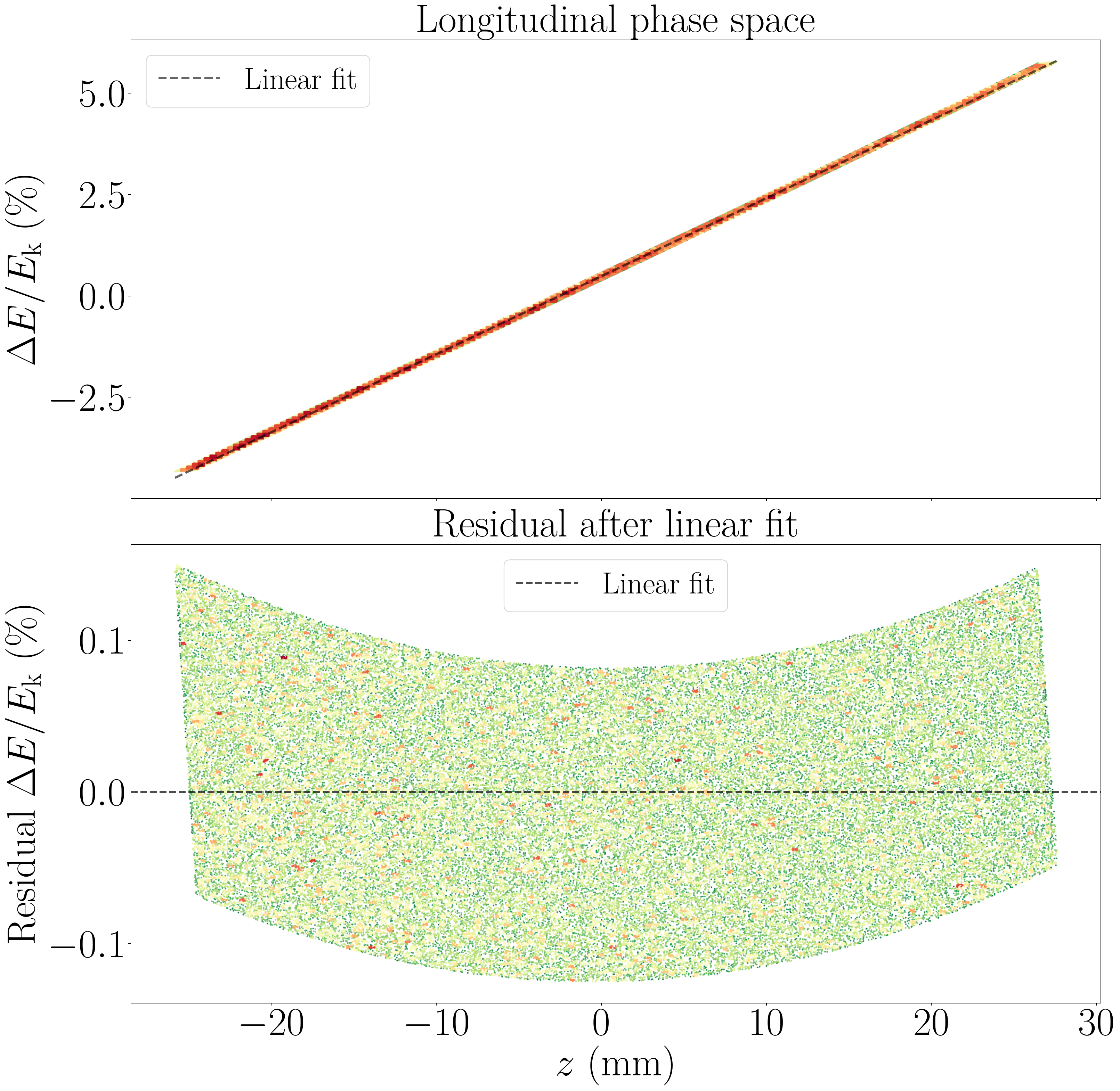}
  \caption{\textbf{Top:} Longitudinal phase space, scaled to ${\Delta E}/{E_{\mathrm{k}}}$, together with a linear fit. The simulated distribution (\SI{11}{\mega \electronvolt}$\pm 5 \; \%$) is shown after capture and before the cavity kick. \textbf{Bottom:} Residual between the phase-space coordinates and the linear fit, showing that although the longitudinal phase space is nearly perfectly correlated, a slight curvature remains.}
  \label{fig_app_curvature_11}
\end{figure}

\begin{figure}[ht]
  \centering
  \includegraphics[width=1\columnwidth]{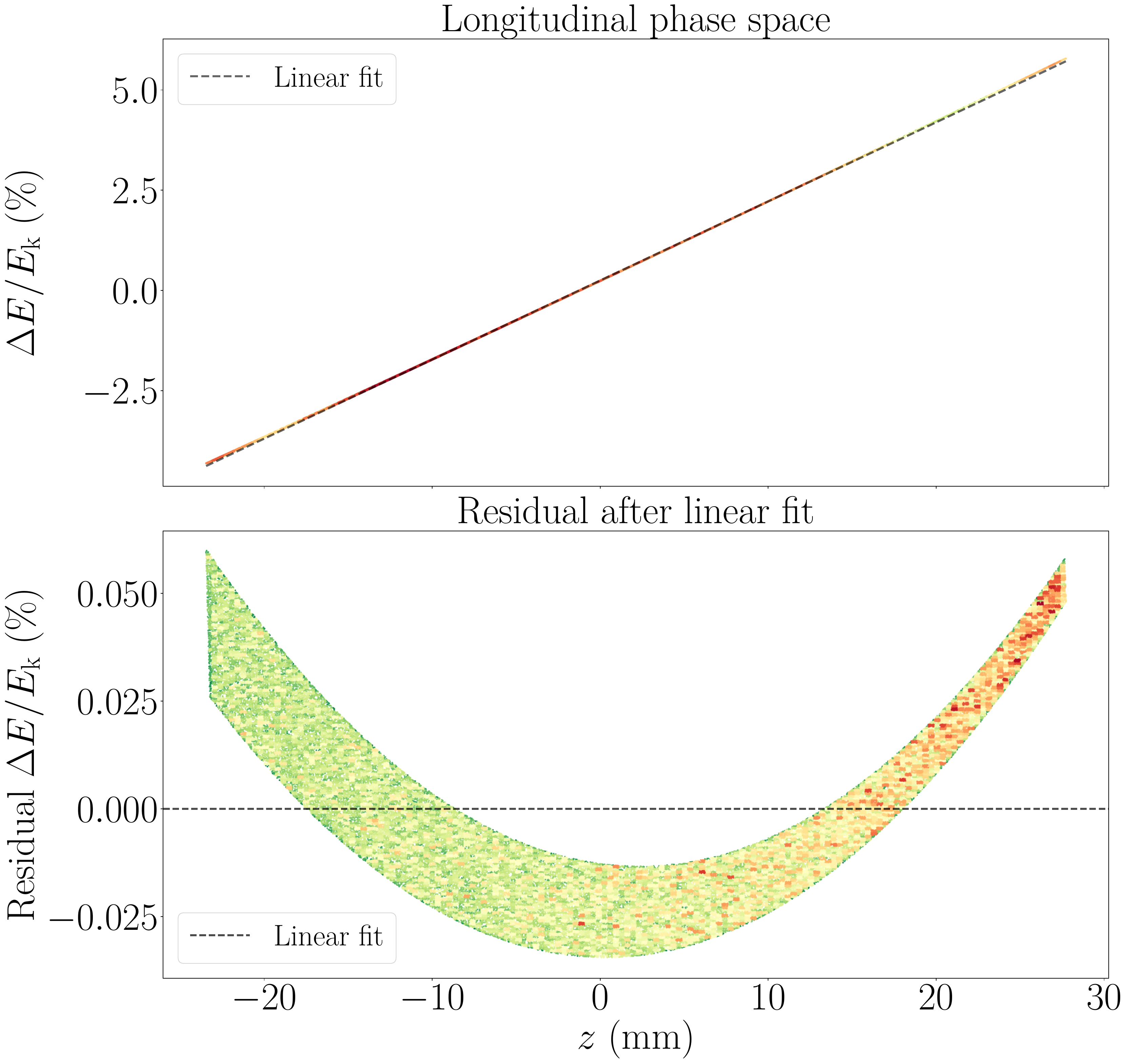}
  \caption{\textbf{Top:} Longitudinal phase space, scaled to ${\Delta E}/{E_{\mathrm{k}}}$  with a linear fit. The simulated distribution (\SI{27}{\mega \electronvolt} $\pm 5 \; \%$) is shown after capture and before the cavity kick. \textbf{Bottom:} Residual between the phase-space coordinates and the linear fit. The asymmetric width results from the energy-dependent divergence of the TNSA beam, since higher-energy particles leave the target with a smaller opening angle.}
  \label{fig_app_curvature_27}
\end{figure}

Figure \ref{fig_app_curvature_11} shows the longitudinal phase space of a TNSA distribution within \SI{11}{\mega \electronvolt} $\pm 5 \; \%$  immediately before the cavity kick. While the phase space itself \textbf{(top)} shows no visible nonlinear features, comparison with a linear fit through the residual $r = y- y_{\mathrm{fit}}$ reveals a measurable curvature. The amplitude of the curvature shown in Fig. \ref{fig_app_curvature_11} \textbf{(bottom)} is determined by the nonlinear kinematics during the drift described above, whereas the width of the distribution results from the initial divergence of the TNSA beam, as already described by Eq. \ref{eq_drift_long_emit}. For comparison, a distribution with an initial energy window of \SI{27}{\mega\electronvolt} $\pm 5\%$ is shown in Fig. \ref{fig_app_curvature_27}.

Since the divergence of TNSA beams decreases with increasing energy, the overall width of the longitudinal phase-space distribution likewise decreases (see Fig. \ref{fig_app_curvature_27}). In the ideal case, the cavity compensates the linear phase-space component, such that only the higher-order contributions remain.

\begin{figure}[ht]
  \centering
  \includegraphics[width=1\columnwidth]{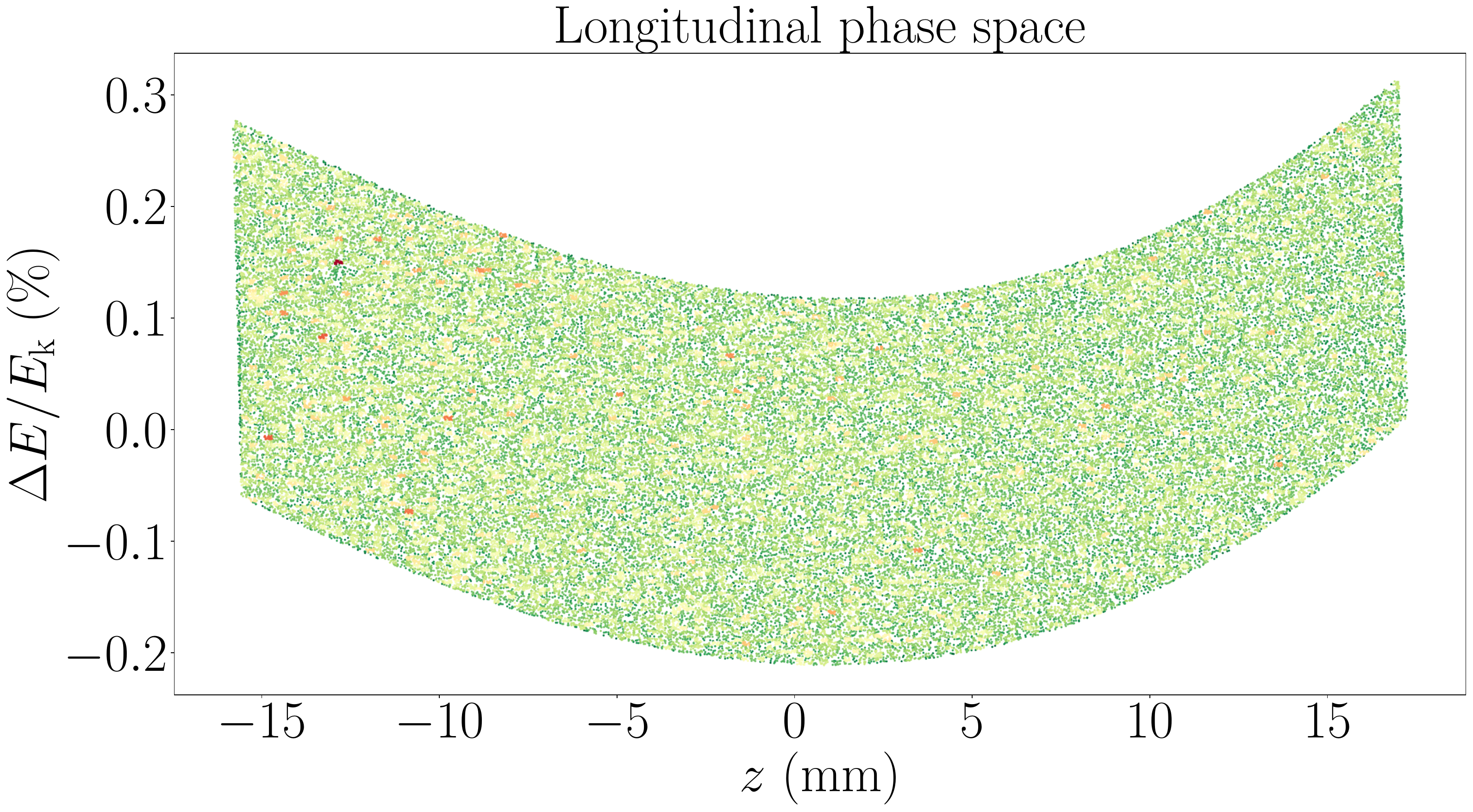}
  \caption{Longitudinal phase space after the cavity kick for a distribution with an initial energy window of \SI{11}{\mega \electronvolt} $\pm 5 \; \%$.}
  \label{fig_app_curvature_kick}
\end{figure}

Figure \ref{fig_app_curvature_kick} illustrates this effect, with the cavity removing the linear component of the longitudinal phase-space distribution such that the remaining structure is nearly identical to the residual shown in Fig. \ref{fig_app_curvature_11}.
\newpage

\bibliography{apssamp}

\end{document}

%% file: results.tex
\newcommand{\varRefEnN}{11.4}
\newcommand{\varRefEn}{\SI{\varRefEnN}{\mega \electronvolt}}
\newcommand{\varEnBndN}{5}
\newcommand{\varEnBnd}{$\pm \varEnBndN \; \%$}

\newcommand{\varEmSolTL}{\SI{13}{\milli \meter \milli \radian}}

\newcommand{\varRefPartNum}{$10^{10}$}

\newcommand{\varPartNumEnAcc}{$8 \times 10^{8}$}
\newcommand{\varPartNumEnAccAPL}{$10^{9}$}
\newcommand{\varEmSolInj}{12}

\newcommand{\varEmAPLInj}{6}

\newcommand{\varPartNumTKAcc}{$2 \times 10^{8}$}
\newcommand{\varPartNumTKAccAPL}{$9 \times 10^{8}$}

\newcommand{\varSpChLim}{$10^{10}$}

%% file: media/lens.tex
\definecolor{amethyst}{rgb}{0.6, 0.4, 0.8}
\definecolor{azure}{rgb}{0.0, 0.5, 1.0}
\definecolor{aqua}{rgb}{0.0, 1.0, 1.0}
\definecolor{darkgreen}{rgb}{0.0, 0.5, 0.0}

\def\wa{4}
\def\we{\wa+\d}   
\def\d{0.96}
\def\R{8}

    \begin{tikzpicture}[scale=0.85]

        \draw[fill=white] (0,-3) rectangle (10,3);
        \draw[dashed] (0.5,0) -- (9.5,0);
        \fill[lightgray] (0.8,-1.5) rectangle (1,1.5);

        \draw[thick] (\we,0) arc (0:20:\R);
        \draw[thick] (\we,0) arc (0:-20:\R);
        \draw[thick] (\wa,0) arc (180:160:\R);
        \draw[thick] (\wa,0) arc (180:200:\R);

        \draw[dashed] (\wa+\d/2,-2.9) -- (\wa+\d/2, 2.9);

        \draw[thick, darkgreen] (1,0) -- (\wa+\d/2, 2.7);
        \draw[thick, darkgreen] (1,0) -- (\wa+\d/2, -2.7);
        \draw[thick, darkgreen] (\wa+\d/2, -2.7) -- (\wa+\d/2+2, -2.7);
        \draw[thick, darkgreen, dashed] (\wa+\d/2+2, -2.7) -- (\wa+\d/2+4, -2.7);
        \draw[thick, darkgreen] (\wa+\d/2, 2.7) -- (\wa+\d/2+2, 2.7);
        \draw[thick, darkgreen, dashed] (\wa+\d/2+2, 2.7) -- (\wa+\d/2+4, 2.7);

        \draw[darkgreen] (1.8,0) arc (0:31:1);

        \draw[thick, <->] (1,-1.5) -- (\wa+\d/2,-1.5);
        \draw[thick, <->] (8,0) -- (8,2.7);

        \node[gray] at (1.6,1.7) {TNSA Target};
        \node[darkgreen, font=\large] at (2.1,0.3) {$\theta_{\mathrm{a}}$};
        \node[font=\large] at (6.1,2.3) {Capture lens};
        \node[font=\large] at (2.1,-1.9) {$d = f$};
        \node[font=\large] at (8.4,1.35) {$R_{\mathrm{a}}$};

    \end{tikzpicture}

    

%% file: media/tab_sis_injection.tex
\begin{tabular}{l >{\raggedright\arraybackslash}p{0.45\linewidth} ll}
\textbf{Variable} & \textbf{Description}                          & \textbf{Value} & \textbf{Unit} \\ \hline
\multicolumn{4}{l}{\textbf{TK/SIS18 acceptance}}                                                     \\ \hline
$\mathbf{E_{k}}$& Reference energy                & \varRefEnN{}        & MeV           \\
$\mathbf{\Delta E/E_{k}}$& Energy spread (SIS18)            & \textpm 0.2      & \%            \\
$\mathbf{\varepsilon_n}$    & Equivalent maximum norm. emittance (TK)            & 3             & mm mrad           \\
\multicolumn{4}{l}{\textbf{Generated beam parameters at TNSA target}}                            \\ \hline
$\mathbf{\theta_{max}}$  & Max. half opening angle      & 26.5             & deg            \\
$\mathbf{\Delta E_{S}/E_{k}}$& Initial energy spread        & $\pm \varEnBndN{}$            & \%            \\
$\mathbf{N_{abs}}$& Initial particles    & \varRefPartNum{}              &              \\
\multicolumn{4}{l}{\textbf{Simulation results - solenoid}}                            \\ \hline
$\mathbf{N_{TK}}$  & \textbf{Particles at TK inj.}      &\cellcolor[HTML]{EFEFEF} \varPartNumEnAcc{}            &           \\
$\mathbf{\varepsilon_{n,TK}}$  & Emittance at TK inj.       & \varEmSolInj{}            & mm mrad            \\
$\mathbf{z_{m,TK}}$  & Long. bunch length       & 15            & mm            \\
$\mathbf{\Delta E_{TK}/E_{k}}$  & Energy spread at TK inj.     & $\pm 0.1$            & \%            \\
$\mathbf{N_{SIS}}$  & \textbf{Particles exiting TK }     &\cellcolor[HTML]{EFEFEF} \varPartNumTKAcc{}           &           \\
$\mathbf{\varepsilon_{n,SIS}}$  & Emittance at TK exit       & 3            & mm mrad            \\
$\mathbf{\Delta E_{SIS}/E_{k}}$  & Energy spread at TK exit      & $\pm 0.05$            & \%            \\
            
\end{tabular}

%% file: media/tab_sis_injection_apl.tex
\begin{tabular}{llll}
\textbf{Variable} & \textbf{Description}                          & \textbf{Value} & \textbf{Unit} \\ \hline

\multicolumn{4}{l}{\textbf{Simulation results - APL}}                            \\ \hline
$\mathbf{N_{TK}}$  & Particles at TK inj.      &\cellcolor[HTML]{EFEFEF} \varPartNumEnAccAPL{}            &           \\
$\mathbf{\varepsilon_{n,TK}}$  & Emittance at TK inj.       &\varEmAPLInj{}   & mm mrad            \\
$\mathbf{z_{m,TK}}$  & Long. bunch length       & 20            & mm            \\
$\mathbf{\Delta E_{TK}/E_{k}}$  & Energy spread at TK inj.    & $\pm 0.2$            & \%            \\
$\mathbf{N_{SIS}}$  & Particles exiting TK      &\cellcolor[HTML]{EFEFEF} \varPartNumTKAccAPL{}            &           \\
$\mathbf{\varepsilon_{n,SIS}}$  & Emittance at TK exit       & 3            & mm mrad            \\
$\mathbf{\Delta E_{SIS}/E_{k}}$  & Energy spread at TK exit      & $\pm 0.2$            & \%            \\
            
\end{tabular}